\documentclass[final]{IEEEtran}
\usepackage{cite}
\usepackage{graphicx}
\usepackage{picinpar}
\usepackage[cmex10]{amsmath}
\usepackage{amsmath,amsfonts,amssymb}
\usepackage{subfigure}

\usepackage{algorithm}
\usepackage{algorithmic}
\usepackage{stfloats}
\usepackage{bm}
\usepackage{color}
\usepackage{stfloats}
\usepackage{url}
\usepackage{makecell}
\usepackage{multirow}
\usepackage{array}
\usepackage{booktabs}

\newcommand{\e}{\begin{equation}}
\newcommand{\ee}{\end{equation}}
\newcommand{\eqn}{\begin{eqnarray}}
\newcommand{\eeqn}{\end{eqnarray}}
\begin{document}

\title{Model-Driven Deep Learning Based Channel Estimation and Feedback for Millimeter-Wave Massive Hybrid MIMO Systems}

\author{Xisuo Ma, Zhen Gao,~\IEEEmembership{Member,~IEEE}, Feifei Gao,~\IEEEmembership{Fellow,~IEEE}, 
and Marco Di Renzo,~\IEEEmembership{Fellow,~IEEE} ~\IEEEmembership{}

\thanks{The work of Z. Gao was supported by the Beijing Municipal Natural Science Foundation under Grant L182024, National Natural Science Foundation of China under Grant 62071044, the Young Elite Scientists Sponsorship Program by CAST, and in part by the Talent Innovation Project of BIT. The work of M. Di Renzo was supported in part by the European Commission through the H2020 ARIADNE project under grant agreement no. 871464 and through the H2020 RISE-6G project under grant agreement no. 101017011. The codes and some other materials about this work may be available at https://gaozhen16.github.io. \it{(Corresponding author: Zhen Gao.)}}
\thanks{X. Ma and Z. Gao are with both Advanced Research Institute of Multidisciplinary Science (ARIMS) and School of Information and Electronics,
Beijing Institute of Technology (BIT), Beijing 100081, China (E-mail: maxisuo@bit.edu.cn, gaozhen16@bit.edu.cn).}
\thanks{F. Gao is with the Institute for Artificial Intelligence, Tsinghua University (THUAI), State Key   Lab of Intelligent Technologier and Systems, Tsinghua University, Beijing National Research Center for Information Science and Technology (BNRist), Department of Automation, Tsinghua University, Beijing 100084, China (E-mail: feifeigao@ieee.org).}
\thanks{M. Di Renzo is with Universit\'e Paris-Saclay, CNRS, CentraleSup\'elec, Laboratoire des Signaux et Syst\`emes, 3 Rue Joliot-Curie, 91192 Gif-sur-Yvette, France (E-mail: marco.di-renzo@universite-paris-saclay.fr).}
}
\maketitle
\begin{abstract}
This paper proposes a model-driven deep learning (MDDL)-based channel estimation and feedback scheme for wideband millimeter-wave (mmWave) massive hybrid multiple-input multiple-output (MIMO) systems, where the angle-delay domain channels' sparsity is exploited for reducing the overhead.
First, we consider the uplink channel estimation for time-division duplexing systems.
To reduce the uplink pilot overhead for estimating high-dimensional channels from a limited number of radio frequency (RF) chains at the base station (BS), 
we propose to jointly train the phase shift network and the channel estimator as an auto-encoder. 
Particularly, by exploiting the channels' structured sparsity from an {\it{{\textbf{a priori}}}} model and learning the integrated trainable parameters from the data samples,
the proposed multiple-measurement-vectors learned approximate message passing (MMV-LAMP) network with the devised redundant dictionary can jointly recover multiple subcarriers' channels with significantly enhanced performance. 
Moreover, we consider the downlink channel estimation and feedback for frequency-division duplexing systems.
Similarly, the pilots at the BS and channel estimator at the users can be jointly trained as an encoder and a decoder, respectively.
Besides, to further reduce the channel feedback overhead, only the received pilots on part of the subcarriers are fed back to the BS, which can exploit the MMV-LAMP network to reconstruct the spatial-frequency channel matrix.
Numerical results show that the proposed MDDL-based channel estimation and feedback scheme outperforms the state-of-the-art approaches.
\end{abstract}

\vspace{-1mm}
\IEEEpeerreviewmaketitle

\section{Introduction}
Millimeter-wave (mmWave) has been widely recognized as a key technology in future wireless communication systems, since the abundant bandwidth resources can significantly increase the throughput \cite{heath_JSTSP, GZ_WC}.
Moreover, to mitigate the severe propagation loss in the mmWave band, massive multiple-input multiple-output (MIMO) is usually adopted to perform beamforming \cite{LULU_JSTSP, CM_beam}.
However, the fully-digital massive MIMO architecture gives rise to an unaffordable hardware cost and power consumption, where a dedicated radio frequency (RF) chain is required for each antenna.
In order to circumvent the technical hurdle and facilitate the deployment of mmWave massive MIMO systems in practice, the phase shift network (PSN) based hybrid MIMO architecture has been widely adopted to achieve a large array gain with a much smaller number of RF chains \cite{YuWei_JSTSP, MJN_WCL, LY_JSTSP}.

To fully capitalize on the large spatial degrees of freedom in mmWave massive MIMO systems, channel state information (CSI) at the base station (BS) is essential,
since beamforming, signal detection, and interference alignment heavily rely on accurate CSI at the BS \cite{SWQ_TVT}.
As for time-division duplexing (TDD) systems, estimating the high-dimensional uplink mmWave massive MIMO channels from limited number of RF chains at the BS suffers from an excessively high pilot overhead \cite{LY_TWC}.
As for frequency-division duplexing (FDD) systems, the downlink high-dimensional channel is first obtained at the users using very few RF chains, and is then fed back to the BS. 
In this case, the prohibitively high channel estimation and feedback overhead problem is more severe~\cite{GZ_TSP}.
\subsection{Related Work}
To this end, by exploiting the sparsity of massive MIMO channels in the angle-domain and/or delay-domain, several low overhead channel estimation and feedback solutions have been proposed \cite{SWQ_TVT, low_frequency_massive_MIMO1, GZ_TSP, LAW_Tcom, LY_TWC, WZW_TVT, LXC_TWC}.
Specifically, by exploiting the temporal correlation of time-varying channels, the authors of \cite{SWQ_TVT} proposed a differential channel estimation and feedback scheme for FDD massive MIMO systems with reduced overhead, and a structured compressive sampling matching pursuit (S-CoSaMP) algorithm to acquire a reliable CSI at the BS.
In \cite{GZ_TSP}, the authors proposed a spatially common sparsity based adaptive channel estimation and feedback scheme for FDD massive MIMO systems, which adapted the training overhead and pilot design to reliably estimate and feed back the downlink CSI with reduced overhead.
Moreover, by introducing an enhanced Newtonized orthogonal matching pursuit (eNOMP) algorithm, the authors of \cite{low_frequency_massive_MIMO1} proposed an efficient downlink channel reconstruction-based transceiver for FDD massive MIMO systems.
However, these schemes \cite{SWQ_TVT, low_frequency_massive_MIMO1, GZ_TSP} were mainly proposed for low-frequency massive MIMO systems using a fully-digital array.

As for the mmWave hybrid MIMO,
by exploiting the channel sparsity in both angle and delay domains, a closed-loop sparse channel estimation scheme for TDD systems was proposed in \cite{LAW_Tcom}, which utilized ESPRIT-type algorithms to acquire super-resolution estimates of the angle of arrivals/departures (AoAs/AoDs) and of the delays of multipath components with low overhead.
In \cite{LY_TWC}, two high-resolution channel estimation schemes based on the ESPRIT algorithm were proposed for broadband mmWave massive MIMO systems. 
By exploiting the mmWave channels' sparsity, the authors of \cite{WZW_TVT} proposed a compressive sensing (CS) greedy algorithm based channel estimation solution for reducing the channel estimation overhead.
Additionally, by exploiting the 3-D clustered structure exhibited in the virtual AoA-AoD-delay domain, an approximate message passing (AMP) with the nearest neighbor pattern learning algorithm was proposed to estimate the broadband mmWave massive MIMO-OFDM channels in \cite{LXC_TWC}.
Although the overhead is reduced, the computational complexity of ESPRIT techniques \cite{LAW_Tcom, LY_TWC} and greedy algorithms \cite{WZW_TVT} can be prohibitively high due to the matrix inversion and singular value decomposition (SVD) operations.
Besides, although AMP algorithms based solutions \cite{LXC_TWC} can reduce the computational complexity, they heavily rely on {\it{a priori}} models, which would lead to performance degradation since {\it{a priori}} models may not always be consistent with the actual systems.
\subsection{Motivations}
Recently, the successful application of deep learning in various fields, particularly in computer science, has gained major attention in the communication community, and has promoted an increasing interest in applying it to address communication and signal processing problems \cite{QZJ_WC, GUI_WC, QZJ_TCCN, Marco_Tcom}.
The deep learning based intelligent communication paradigm has attained manifold accomplishments, including channel coding \cite{decoder_design}, random access \cite{random_access}, beamforming design \cite{beam_design1,HHJ_hybrid_precoding, beamforming_ICC, YuWei_Globecom, YuWei_TWC}
activity and signal detection \cite{activity_detection, signal_detection}, autoencoder-based end-to-end communication system \cite{autoencoder_system}, CSI feedback \cite{CSI_feedback1, CSI_feedback2, CSI_feedback3}, and channel estimation \cite{LY_JSTSP, channel_estimation1, MXS_TVT}, etc.
To be specific, pure data-driven deep learning based solutions often employ deep neural networks (DNNs),  including fully-connected neural networks and/or convolutional neural networks (CNNs), as a black box to design communication signal processing modules without any {\it{a priori}} model information.
Moreover, a large amount of training data samples are required to optimize the neural network through a customized loss function and learning strategy.

In particular, in order to overcome the high-computational complexity and fully exploit the spatial information, the authors of \cite{HHJ_hybrid_precoding} proposed a deep-learning-enabled mmWave massive MIMO framework for effective hybrid precoding, in which each selection of precoders for obtaining the optimized decoder is regarded as a mapping relation in the DNN.
In addition, the authors of \cite{YuWei_Globecom} proposed a DNN-based approach for channel sensing and downlink hybrid analog-digital beamforming, which was generalizable for any numbers of users by decomposing the deep learning architecture into multiple parallel independent single-user DNNs.
By considering the multiuser channel estimation and feedback problem as a distributed source coding problem, the authors of \cite{YuWei_TWC} proposed a joint design of pilots and a novel DNN architecture, which mapped the feedback bits from all the users directly into the precoding matrix at the BS.
By exploiting an unsupervised maching learning (ML) model, i.e., an autoencoder, the authors of \cite{beamforming_ICC} presented a linear autoencoder-based beamformer and combiner design, which maximizes the achievable rates over a mmWave channel.
Moreover, in order to address the overwhelming feedback overhead of FDD massive MIMO systems, the authors of \cite{CSI_feedback1} proposed a CS-ReNet framework, where the CSI was first compressed at the users based on CS methods and then reconstructed at the BS using a deep learning-based recovery solver.
However, in practical scenarios, there exists various interference and non-linear effects.
Therefore, the authors of \cite{CSI_feedback3} designed a deep learning-based denoising network, called DNNet, to improve the performance and robustness of channel feedback.
Additionally, by exploiting the spatial, temporal, and frequency correlations, the authors of \cite{LY_JSTSP} employed a CNN to address the channel estimation problem for mmWave massive hybrid MIMO systems.
However, the performance of those data-driven approaches heavily depends on the quantity and quality of the training data samples, but good data sets are usually difficult to be obtained in practice and parctical issues such as training over-fitting could degrade the capability of the system to generalize.
In addition, data-driven approaches lack interpretability and trustability that are major strengths of model-driven signal processing.
Moreover, compared with conventional model-based methods, model-driven approaches have better denoising capabilities by utilizing the powerful data processing capabilities and denoising capabilities of neural networks.

Therefore, different from pure data-driven and conventional model-based approaches, model-driven deep learning (MDDL)-based approaches construct the network structure by exploiting {\it{a priori}} knowledge from known physical mechanisms, such as well-developed channel models and transmission protocols. 
Note that the MDDL-based approaches retain some of the advantages of conventional model-based iterative methods, which includes exploiting some {\it a priori} information to be trained with fewer trainable parameters and with less data samples, e.g., the structured sparsity of the mmWave channels can be exploited. 
Moreover, they can further retain the learning ability of deep learning methods and avoid the performance degradation caused by the mismatch between the predetermined parameters (based on an assumed model) and the true optimal parameters (based on empirical data samples).
By leveraging some {\it{a priori}} information, model-driven methods require fewer parameters to be learned and less samples for training as compared to pure data-driven deep learning solutions \cite{Marco_VTM, channel_estimation2, LAMP_TSP, HHT_WC, AMP_WCL}.
Specifically, the authors of \cite{channel_estimation2} proposed an MDDL-based downlink channel reconstruction scheme for FDD massive MIMO systems, where a powerful neural network, named You Only Look Once (YOLO), was introduced to enable a rapid estimation process of the model parameters.
Moreover, \cite{AMP_WCL} proposed a novel AMP-based network with deep residual learning, referred to as LampResNet, to estimate the beamspace channel for mmWave massive MIMO systems.
\subsection{Our Contributions}
This paper proposes an MDDL-based channel estimation and feedback scheme for wideband mmWave massive hybrid MIMO systems, where  the angle-delay domain channels' sparsity is exploited for reducing the overhead.
First, we consider the uplink channel estimation for TDD systems.
To reduce the uplink pilot overhead for estimating the high-dimensional channels from a limited number of RF chains at the BS,
we propose to jointly train the PSN and the channel estimator as an auto-encoder.
Particularly, by learning the integrated trainable parameters from data samples and exploiting the channels' structured sparsity from an {\it{a priori}} model,
the proposed multiple-measurement-vectors learned approximate message passing (MMV-LAMP) network with the devised redundant dictionary can jointly recover multiple subcarriers' channels with significantly enhanced performance. 
Moreover, we consider the downlink channel estimation and feedback for FDD systems.
Similarly, the pilots at the BS and channel estimator at the users can be jointly trained as an encoder and a decoder, respectively.
Besides, to further reduce the channel feedback overhead, only the received pilots on part of the subcarriers are fed back to the BS, which can exploit the MMV-LAMP network to reconstruct the spatial-frequency channel matrix.
Simulations are conducted to demonstrate the effectiveness of the proposed MDDL-based channel estimation and feedback scheme over the conventional approaches.

The main contributions of this paper are summarized as follows:
\begin{itemize}
	\item Operations in the complex domain are well supported by most deep learning frameworks.
	However, the beamforming/combining matrix is a complex-valued matrix and satisfies the constant modulus constraint due to the RF PSN adopted in the hybrid MIMO architecture.
	To this end, we design a novel fully-connected channel compression network (CCN) as the encoder to compress the high-dimensional channels, and the network parameters are defined as the real-valued phases of the PSN (i.e., beamforming/combining matrix in channel estimation).
	\item To reliably reconstruct the channels from the compressed measurements, 
	we propose a channel reconstruction network (CRN) based on a developed MMV-LAMP network with the devised redundant dictionary as the decoder, which can exploit the {\it{a priori}} model and learn the optimal parameters from data to jointly recover multiple subcarriers' channels with significantly enhanced performance.
	\item To effectively estimate the channels from compressed feedback signals, a feedback based channel reconstruction network (FCRN) is proposed. The FCRN consists of a feedback reconstruction sub-network (FRSN) and a CRN. The FRSN is based on the MMV-LAMP network and can exploit the delay-domain sparsity of the channels to reliably reconstruct the compressed channel.
	\item Due to the mismatch between continuous AoAs/AoDs and the limited angle resolution of spatial-angular transform matrix, the resulted power leakage can weaken the channel sparsity represented in the angle domain. Hence, by quantizing the angles with a finer resolution, we design a redundant dictionary to further improve the sparse channel estimation performance.
	\item To evaluate the superiority of the proposed solution that jointly trains the pilots and channel estimator, we further consider scenarios with fixed scattering environments.
	Simulation results verify that, by learning the characteristics of the data samples with fixed scattering, the optimized CCN and MMV-LAMP network can well match the channel environments with improved performance.
\end{itemize}

\textit{Notations}: Throughout this paper, scalar variables are denoted by normal-face letters, while boldface lower and upper-case symbols denote column vectors and matrices, respectively. 
Superscripts ${(  \cdot  )^{\text{T}}}$, ${(  \cdot  )^ * }$ and ${(  \cdot  )^{\text{H}}}$ denote the transpose, conjugate and Hermitian transpose operators, respectively. 
${\left\| {\mathbf{a}} \right\|_0}$ and ${\left\| {\mathbf{A}} \right\|_F}$ denote the ${\ell}_{0}$-norm of ${\mathbf{a}}$ and the Frobenius norm of ${\mathbf{A}}$, respectively.
${\left[ {\mathbf{a}} \right]_m}$ and ${\left[ {\mathbf{A}} \right]_{m,n}}$ are the $m$-th element of ${\mathbf{a}}$ and the $m$-th row and the $n$-th colomn element of ${\mathbf{A}}$, respectively. 
${\mathbf{A}}( {m,:} )$ and ${\mathbf{A}}( {:,n} )$ denote the $m$-th row vector and the $n$-th colomn vector of ${\mathbf{A}}$, respectively.
${\left. {\mathbf{A}} \right|_{\mathbf{\Omega }}}$ denotes a sub-matrix by selecting the rows of ${\mathbf{A}}$ according to the ordered set ${\bm{\Omega}}$ and ${\left\{ {\bm{\Omega }} \right\}_m}$ is the $m$-th element of the set ${\bm{\Omega }}$.
${{\mathbf{e}}^{{\text{j}}\left[ {\bm{\Xi }} \right]}}$ denotes a complex matrix with its element being ${\left[ {{{\mathbf{e}}^{{\text{j}}\left[ {\bm{\Xi }} \right]}}} \right]_{m,n}} = {e^{{\text{j}}{{\left[ {\bm{\Xi }} \right]}_{m,n}}}}$, and ${\bm{\Xi }}$ is a real matrix.
Finally, $\partial (  \cdot  )$ is the first-order partial derivative operation.

\section{System Model}
Consider a mmWave massive MIMO system with hybrid beamforming,
where the BS is equipped with a uniform linear array (ULA) and comprises ${N_{{\rm{BS}}}}$ antennas
and ${N_{{\rm{RF}}}}$ RF chains, and the $U$ users have a single-antenna.
At the BS, the PSN is employed to connect a large number of antennas with a much fewer number of RF chains (i.e., ${{N_{{\text{BS}}}} \gg {N_{{\text{RF}}}}}$), and orthogonal frequency division multiplexing (OFDM) with $K$ subcarriers is adopted to combat the frequency selective fading of the mmWave channels.

\subsection{Uplink Channel Estimation for TDD Systems}
Firstly, we consider the uplink channel estimation for TDD systems.
The uplink channel estimation stage includes $Q$ OFDM symbols (i.e., $Q$ time slots) dedicated for channel estimation.
For a certain user$\footnote{Consider the uplink multi-user channel estimation, if $U$ users adopt mutually orthogonal pilot signals, the pilot signals associated with different users can be distinguished and then respectively processed.}$,
in order to estimate the $k$-th subcarrier's channel, the received baseband signal vector ${{\mathbf{y}}'_{{\text{UL}}}}\left[ {k,q} \right] \in {\mathbb{C}^{{N_{{\text{RF}}}} \times 1}}$  at the BS in the $q$-th time slot can be expressed as
\begin{equation}\label{uplink received signal q}
{{{\mathbf{y}}}'_{{\text{UL}}}}\left[ {k,q} \right] = {\mathbf{F}}_{{\text{UL}}}^{\text{H}}\left[ q \right]{{\mathbf{h}}_{{\text{UL}}}}\left[ k \right]x\left[ {k,q} \right] + {{{\mathbf{\bar n'}}}_{{\text{UL}}}}\left[ {k,q} \right],
\end{equation}
where $1 \le q \le Q$, $1 \le k \le K$,
${{\mathbf{F}}_{{\text{UL}}}}\left[ {q} \right] \in {\mathbb{C}^{{N_{{\text{BS}}}} \times {N_{{\text{RF}}}}}}$ denotes the uplink combining matrix at the BS,
${{\mathbf{h}}_{{\text{UL}}}}\left[ k \right] \in {\mathbb{C}^{{N_{{\text{BS}}}} \times 1}}$ is the uplink $k$-th subcarrier channel,
$x\left[ {k,q} \right] \in \mathbb{C}$ is the transmitted pilot symbol, 
and ${{{\mathbf{\bar n'}}}_{{\text{UL}}}}\left[ {k,q} \right] \sim \mathcal{C}\mathcal{N}( {0,\sigma _n^2{\mathbf{I}}_{N_{\text{RF}}}} )$ is the effective noise modeled at the receiver (front-end) level.

Then, the received baseband signal is post-processed by multiplying it by ${x^ * }\left[ {k,q} \right]$, i.e.,
\begin{align}\label{received_signal_tran}
{{\mathbf{y}}_{{\text{UL}}}}\left[ k,q \right] = {{\mathbf{y}}'_{{\text{UL}}}}\left[ k,q \right]{x^ * }\left[ {k,q} \right] 
	= {\mathbf{F}}_{{\text{UL}}}^{\text{H}}\left[ q \right]{{\mathbf{h}}_{{\text{UL}}}}\left[ k \right] + {{{ {\mathbf{n}}}}_{{\text{UL}}}}\left[ {k,q} \right],
\end{align}
where we assume that $x\left[ {k,q} \right]{x^ * }\left[ {k,q} \right] = 1$ 
and ${{{ {\mathbf{n}}}}_{{\text{UL}}}}\left[ {k,q} \right] = {{{\bar {\mathbf{n}}}'}_{{\text{UL}}}}\left[ {k,q} \right]{x^ * }\left[ {k,q} \right]$.
Note that, due to the constant modulus constraint of the adopted fully-connected RF PSN at the BS, 
the uplink combining matrix ${{\mathbf{F}}_{{\text{UL}}}}\left[ {q} \right]$, $\forall q$, can be expressed as
${\big[ {{{\mathbf{F}}_{{\text{UL}}}}\left[ q \right]} \big]_{m,n}} = \frac{1}{{\sqrt {{N_{{\text{BS}}}}} }}{e^{{\text{j}}{{\left[ {{{\bm{\Xi }}_{{\text{UL}}}}} \right]}_{m,n}}}}$ for $1 \le m \le N_{\text{BS}}$, $1 \le n \le N_{\text{RF}}$,
and $\left[{{{{{\bm{\Xi }}_{{\text{UL}}}}}}}\right]_{m,n}$ denotes the phase value connecting the $m$-th antenna and the $n$-th RF chain$\footnote{Note that changing the phase values of the PSN does not require a re-synchronization of the whole system. This is because: i) The phase values of the phase shifts will be changed in the guard interval before each pilot OFDM symbol; ii) The synchronization of frame and symbol can be obtained based on the preambles transmitted before the pilot symbols; iii) When to adjust the phase shifts can be exactly calculated according to the synchronization information and the predefined signal frame structure, and the adjustment of the phase shifts can be controlled according to the system clock.}$.
By collecting ${{\mathbf{y}}_{{\text{UL}}}}\left[ {k,q} \right]$ for $1 \le q \le Q$ together,
the aggregate received signals ${{\mathbf{y}}_{{\text{UL}}}}\left[ k \right] \in {\mathbb{C}^{M \times 1}}$ ($M = Q{N_{{\text{RF}}}}$) can be written as
\begin{equation}\label{uplink received signal Q}
	{{\mathbf{y}}_{{\text{UL}}}}\left[ k \right] = {\mathbf{F}}_{{\text{UL}}}^{\text{H}}{{\mathbf{h}}_{{\text{UL}}}}\left[ k \right] + {{{ {\mathbf{n}}}}_{{\text{UL}}}}\left[ {k} \right],	
\end{equation}
where ${{\mathbf{y}}_{{\text{UL}}}}\left[ k \right] = {\big[ {{\mathbf{y}}_{{\text{UL}}}^{\text{T}}\left[ {k,1} \right], \cdots ,{\mathbf{y}}_{{\text{UL}}}^{\text{T}}\left[ {k,Q} \right]} \big]^{\text{T}}}$,
${{\mathbf{F}}_{{\text{UL}}}} = \big[ {{{\mathbf{F}}_{{\text{UL}}}}\left[ {1} \right], \cdots ,{{\mathbf{F}}_{{\text{UL}}}}\left[ {Q} \right]} \big] \in {\mathbb{C}^{{N_{{\text{BS}}}} \times M}}$,
and ${{{ {\mathbf{n}}}}_{{\text{UL}}}}\left[ k \right] = {\big[ {{ {\mathbf{n}}}_{{\text{UL}}}^{\text{T}}\left[ {k,1} \right], \cdots ,{ {\mathbf{n}}}_{{\text{UL}}}^{\text{T}}\left[ {k,Q} \right]} \big]^{\text{T}}} \in {\mathbb{C}^{M \times 1}}$.
Finally, by stacking ${{\mathbf{y}}_{{\text{UL}}}}\left[ k \right]$ from all subcarriers, the received signals ${{\mathbf{y}}_{{\text{UL}}}}\left[ k \right]$ for $1 \le k \le K$ can be further expressed as
\begin{equation}\label{uplink received signal K}
	{{\mathbf{Y}}_{{\text{UL}}}} = {\mathbf{F}}_{{\text{UL}}}^{\text{H}}{\mathbf{H}}_{{\text{UL}}}^{{\text{sf}}} + { {\mathbf{N}}}_{\text{UL}},
\end{equation}
where ${{\mathbf{Y}}_{{\text{UL}}}} = \big[ {{{\mathbf{y}}_{{\text{UL}}}}\left[ 1 \right], \cdots ,{{\mathbf{y}}_{{\text{UL}}}}\left[ K \right]} \big] \in {\mathbb{C}^{M \times K}}$,
${\mathbf{H}}_{{\text{UL}}}^{{\text{sf}}} = \big[ {{{\mathbf{h}}_{{\text{UL}}}}\left[ 1 \right],\cdots ,{{\mathbf{h}}_{{\text{UL}}}}\left[ K \right]} \big] \in {\mathbb{C}^{{N_{{\text{BS}}}} \times K}}$ denotes the uplink spatial-frequency domain channel matrix,
and ${{{ {\mathbf{N}}}}_{{\text{UL}}}} = \big[ {{{ {\mathbf{n}}}}_{{\text{UL}}}}\left[ 1 \right], \cdots ,{{{{\mathbf{n}}}}_{{\text{UL}}}}\left[ K \right] \big] \in {\mathbb{C}^{M \times K}}$.

\subsection{Downlink Channel Estimation and Feedback for FDD Systems}
Moreover, we consider the downlink channel estimation and feedback for FDD systems.
Specifically, the downlink pilot signals transmitted by the BS can be denoted as ${{\mathbf{f}}_{{\text{DL}}}}\left[ {q} \right]{{s}}\left[ {k,q} \right] \in {\mathbb{C}^{{N_{{\text{BS}}}} \times 1}}$ for $1 \le q \le Q$,
where ${{\mathbf{f}}_{{\text{DL}}}}\left[ {q} \right]$ is the RF pilot signal
and ${{s}}\left[ {k,q} \right]$ is the baseband pilot signal.
Mathematically, the received signal in the $q$-th time slot associated with the $k$-th subcarrier at the user can be written as
\begin{equation}\label{received_signal_1}
	{{y}'_{{\text{DL}}}}\left[ {k,q} \right] = {\mathbf{h}}_{{\text{DL}}}^{\text{T}}\left[ k \right]{{\mathbf{f}}_{{\text{DL}}}}\left[ {q} \right]s\left[ {k,q} \right] + {{\bar {n}}_{{\text{DL}}}}\left[ {k,q} \right],
\end{equation}
where ${{\mathbf{h}}_{{\text{DL}}}}\left[ k \right] \in {\mathbb{C}^{{N_{{\text{BS}}}} \times 1}}$ is the downlink $k$-th subcarrier's channel,
and ${{{\bar {n}}}_{{\text{DL}}}}\left[ {k,q} \right]$
is the complex noise.
Similar to (\ref{received_signal_tran}), the received signal can be further post-processed to obtain
\begin{align}\label{down_received_signal_tran}
{y_{{\text{DL}}}}\left[ {k,q} \right] = {y_{{\text{DL}}}'}\left[ {k,q} \right]{s^ * }\left[ {k,q} \right] 
	= {\mathbf{h}}_{{\text{DL}}}^{\text{T}}\left[ k \right]{{\mathbf{f}}_{{\text{DL}}}}\left[ {q} \right] + {{{n}}_{{\text{DL}}}}\left[ {k,q} \right],
\end{align}
where we assume that $s\left[ {k,q} \right]{s^ * }\left[ {k,q} \right] = 1$
and ${{{n}}_{{\text{DL}}}}\left[ {k,q} \right] = {{\bar{n}}_{{\text{DL}}}}\left[ {k,q} \right]{s^ * }\left[ {k,q} \right]$.
Similarly, due to the constant modulus constraint of the adopted RF PSN, the RF pilot signal ${{\mathbf{f}}_{{\text{DL}}}}\left[ {q} \right]$, $\forall q$, can be expressed as
${\big[ {{{\mathbf{f}}_{{\text{DL}}}}\left[ q \right]} \big]_{m}} = \frac{1}{{\sqrt {{N_{{\text{BS}}}}} }}{e^{{\text{j}}{{\left[ {{{\bm{\Xi }}_{{\text{DL}}}}} \right]}_{m}}}}$ for $1 \le m \le N_{\text{BS}}$,
and $\left[{{{{{\bm{\Xi }}_{{\text{DL}}}}}}}\right]_{m}$ denotes the phase value connecting the $m$-th antenna and the activated RF chain.
By collecting the received signals from $Q$ time slots, the aggregate received signals can be expressed as
\begin{equation}\label{received_signal_rewritten}
	{{\mathbf{y}}_{{\text{DL}}}}\left[ k \right] = {\mathbf{F}}_{{\text{DL}}}^{\text{T}}{{\mathbf{h}}_{{\text{DL}}}}\left[ k \right] + {{ {\mathbf{n}}}_{{\text{DL}}}}\left[ k \right],
\end{equation}
where ${{\mathbf{y}}_{{\text{DL}}}}\left[ k \right] = {\big[ {{y_{{\text{DL}}}}\left[ {k,1} \right],\cdots ,{y_{{\text{DL}}}}\left[ {k,Q} \right]} \big]^{\text{T}}} \in {\mathbb{C}^{Q \times 1}}$,
${{\mathbf{F}}_{{\text{DL}}}} = \big[ {{{\mathbf{f}}_{{\text{DL}}}}\left[ {1} \right], \cdots ,{{\mathbf{f}}_{{\text{DL}}}}\left[ {Q} \right]} \big] \in {\mathbb{C}^{{N_{{\text{BS}}}} \times Q}}$, and
${{ {\mathbf{n}}}_{{\text{DL}}}}\left[ k \right] = {\big[ {{{{n}}_{{\text{DL}}}}\left[ {k,1} \right],\cdots ,{{{n}}_{{\text{DL}}}}\left[ {k,Q} \right]} \big]^{\text{T}}} \in {\mathbb{C}^{Q \times 1}}$.
Similar to (\ref{uplink received signal K}), we can collect ${{\mathbf{y}}_{{\text{DL}}}}\left[ k \right]$ for $1 \le k \le K$ from $K$ subcarriers to obtain
\begin{equation}\label{downlink received signal K}
	{{\mathbf{Y}}_{{\text{DL}}}} = {\mathbf{F}}_{{\text{DL}}}^{\text{T}}{\mathbf{H}}_{{\text{DL}}}^{{\text{sf}}} + {{ {\mathbf{N}}}_{{\text{DL}}}},
\end{equation}
where ${\mathbf{H}}_{{\text{DL}}}^{{\text{sf}}} = \big[ {{{\mathbf{h}}_{{\text{DL}}}}\left[ 1 \right], \cdots ,{{\mathbf{h}}_{{\text{DL}}}}\left[ K \right]} \big] \in {\mathbb{C}^{{N_{{\text{BS}}}} \times K}}$ denotes the downlink spatial-frequency domain channel matrix,
${{\mathbf{Y}}_{{\text{DL}}}} = \big[ {{{\mathbf{y}}_{{\text{DL}}}}\left[ 1 \right], \cdots ,{{\mathbf{y}}_{{\text{DL}}}}\left[ K \right]} \big] \in {\mathbb{C}^{Q \times K}}$,
and
${{ {\mathbf{N}}}_{{\text{DL}}}} = \big[ {{{ {\mathbf{n}}}_{{\text{DL}}}}\left[ 1 \right], \cdots ,{{ {\mathbf{n}}}_{{\text{DL}}}}\left[ K \right]} \big] \in {\mathbb{C}^{Q \times K}}$.

\subsection{Channel Model}
According to typical mmWave channel models \cite{heath_JSTSP, LY_JSTSP, LY_TWC, GZ_TSP, LAW_Tcom},
the downlink delay-domain continuous channel vector ${\mathbf{h}}_{\text{DL}}( \tau  ) \in {\mathbb{C}^{{N_{{\text{BS}}}} \times 1}}$ can be expressed as
\begin{equation}\label{delay channel}
{\mathbf{h}}_{\text{DL}}( \tau  ) = \sqrt {\frac{{{N_{{\text{BS}}}}}}{L}} \sum\limits_{l = 1}^L {{\beta _l}p( {\tau  - {\tau _l}} ){\mathbf{a}}( {{\varphi _l}} )} ,
\end{equation}
where ${\beta _l} \sim \mathcal{C}\mathcal{N}( {0,\sigma _\alpha ^2} )$ and ${\tau _l}$ denote the propagation gain and delay corresponding to the $l$-th path, respectively,
$p( \tau  )$ is the pulse shaping filter,
and ${{\varphi _l}}$ is the angle-of-departure (AoD) of the $l$-th path at the BS.
Moreover, the frequency-domain channel ${\mathbf{h}}_{\text{DL}}\left[ k \right]$ at the $k$-th subcarrier can be expressed as
\begin{equation}\label{frequency channel}
{\mathbf{h}}_{\text{DL}}\left[ k \right] = \sqrt {\frac{{{N_{{\text{BS}}}}}}{L}} \sum\limits_{l = 1}^L {{\beta _l}{e^{ - {\text{j}}\frac{{2\pi k{f_s}{\tau _l}}}{K}}}{\mathbf{a}}( {{\varphi _l}} )},
\end{equation}
where ${{f_s}}$ is the system sampling rate.

Since the BS is equipped with an ULA, the corresponding array steering vector ${\mathbf{a}}( \theta  ) \in {\mathbb{C}^{{N_{{\text{BS}}}} \times 1}}$ can be written as
\begin{equation}\label{steeting vector}
{\mathbf{a}}( \theta  ) = \frac{1}{{\sqrt {{N_{{\text{BS}}}}} }}{\left[ {1,{e^{ - {\text{j}}\frac{{2\pi d}}{\lambda }\sin ( \theta  )}}, \cdots ,{e^{ - {\text{j}}\frac{{2\pi d}}{\lambda }( {{N_{{\text{BS}}}} - 1} )\sin ( \theta  )}}} \right]^{\text{T}}},
\end{equation}
where $\lambda$ is the carrier wavelength, and $d$ is the adjacent antenna spacing usually satisfying $d = \lambda / 2$.

\section{MDDL-Based TDD Uplink Channel Estimation}
In this section, we first propose an improved frame structure design for optimizing the channel estimation duration.
Secondly, we propose to jointly train the RF PSN and the channel estimator as an auto-encoder.
Finally, we develop an MMV-LAMP network, which can both exploit the structured sparsity from an {\it{a priori}} model and adaptively learn the trainable parameters from the data samples.

\subsection{The Proposed Transmit Frame Structure Design}
\begin{figure}[t]
	\centering
	\includegraphics[width=252pt]{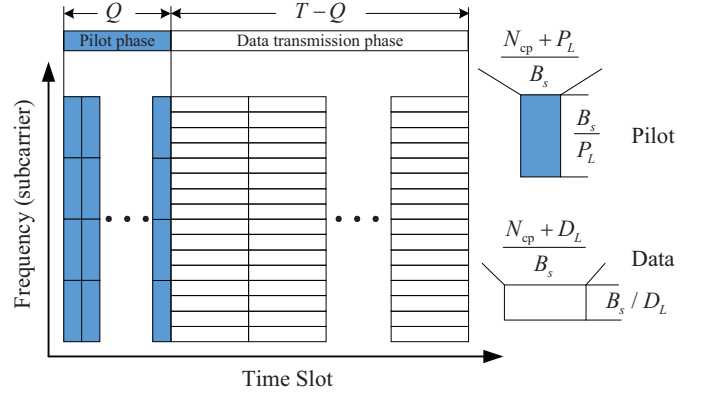}
	\caption{The proposed frame structure for the communication transmission.}
	\label{frame structure}
\end{figure}

The proposed frame structure is illustrated in Fig. \ref{frame structure}, where the cyclic prefix (CP)-OFDM is employed to combat the time dispersive channels and the time-frequency radio resources can be divided into multiple resource elements to convey the pilot signals and payload data. Specifically, a frame comprising $T$ time slots is divided into two phases in the time domain, where the first $Q$ time slots (i.e., pilot phase) are used to transmit pilot signals and the remaining $(T-Q )$ time slots (i.e., data transmission phase) are reserved only for payload data transmission. In the pilot phase, we denote the OFDM's DFT length as $P_{L} = {N_{{\text{cp}}}}$, where ${N_{{\text{cp}}}}$ is the length of CP. Therefore, the subcarrier spacing is ${B_s}/{P_L}$ and each CP-OFDM symbol duration is $( {{N_{{\text{cp}}}} + {P_L}} )/{B_s}$, where ${B_s}$ is the system bandwidth. On the other hand, in the data transmission phase, we consider the OFDM symbol's DFT length is ${D_L} \gg {P_L}$ and thus each CP-OFDM symbol duration is $( {{N_{{\text{cp}}}} + {D_L}} )/{B_s}$.

\subsection{The Developed MMV-LAMP Network}
In this section, we will detail the developed MMV-LAMP network.
Without loss of generality, we consider a typical MMV CS problem
\begin{equation}\label{MMV_problem}
	{\mathbf{Y}} = {\mathbf{AX}} + {\mathbf{N}},
\end{equation}
where ${\mathbf{Y}} \in {\mathbb{C}^{M \times K}}$ is a noisy measurement,
${\mathbf{A}} \in {\mathbb{C}^{M \times N}}$ is a measurement matrix,
${\mathbf{X}}\in {\mathbb{C}^{N \times K}}$ is a sparse matrix whose columns $\big\{ {{\mathbf{X}}( {:,i} )} \big\}_{i = 1}^K$ share a common sparsity, and ${\mathbf{N}} \in {\mathbb{C}^{M \times K}}$ is the additive white Gaussian noise (AWGN).

To solve the MMV CS problem in (\ref{MMV_problem}) efficiently, the developed MMV-LAMP network has two features: i) it fully exploits an {\it{a priori}} model, i.e., the structured sparsity of ${\mathbf{X}}$; ii) by integrating the trainable parameters into the unfolded iterations of conventional AMP algorithms, it can adaptively learn and optimize the network from data samples.
Specifically, as for the $t$-th layer $( {1 \le t \le T} )$ of the developed MMV-LAMP network, the key procedure includes
\begin{subequations}
	\begin{align}
		{{\mathbf{R}}_t} &= {\mathbf{\widehat X}}_{t - 1} + {{\mathbf{B}}}{{\mathbf{V}}_{t-1}} , \\
		{\mathbf{\widehat X}}_t &= {{\eta }}( {{{\mathbf{R}}_t};{{\bm{\theta }}},{\sigma _t}} ), \\
		{{\mathbf{V}}_t} &= {\mathbf{Y}} - {\mathbf{A\widehat X}}_{t} + {b_t}{{\mathbf{V}}_{t - 1}}, \label{residual}
	\end{align}
\end{subequations}
where ${{\mathbf{V}}_0} = \mathbf{Y}$, ${\mathbf{\widehat X}}_0 = \mathbf{0}$, and
\begin{align}
	{\sigma _t} &= \frac{1}{{\sqrt {MK} }}{\left\| {{{\mathbf{V}}_{t - 1}}} \right\|_F}, \label{epsilon_t} \\
	{b_t}{\mathbf{I}} &= \frac{1}{M}\sum\limits_{j = 1}^N {\frac{{\partial {{\left[ {{{\eta}} ( {{{\mathbf{R}}_t};{{\bm{\theta}}},{\sigma _t}} )} \right]}_j}}}{\partial \left[ {{{\mathbf{R}}_t}( {j,:} )} \right]}}. \label{b_t}	
\end{align}
Note that, the residual ${{\mathbf{V}}_t}$ in (\ref{residual}) includes the ``Onsager correction" term ${b_{t}}{{\mathbf{V}}_{t - 1}}$, which is introduced into the conventional AMP algorithms to accelerate the convergence \cite{LAMP_TSP}.
Moreover, the shrinkage function $\eta ( { \cdot ; \cdot } )$ can be expressed as
\begin{equation}\label{eta}
	{\left[ {\eta ( {{{\mathbf{R}}_t};{{\bm{\theta }}},{\sigma _t}} )} \right]_j} = \frac{{{{\mathbf{r}}_{t,j}}}}{{{\pi _t}\left[ {1 + \exp ( {{\psi _t} - \frac{{{\mathbf{r}}_{t,j}^{\text{H}}{{\mathbf{r}}_{t,j}}}}{{2\sigma _t^2{\pi _t}}}})} \right]}},
\end{equation}
where ${{\mathbf{r}}_{t,j}} = {{{\mathbf{R}}_t}( {j,:} )}$ denotes the $j$-th row of the ${{{\mathbf{R}}_t}}$, ${\pi _t}$ and ${\psi _t}$ are respectively given by
\begin{align}
	{\pi _t} &= 1 + \frac{{\sigma _t^2}}{{{\theta _{1}}}}, \label{pi} \\
	{\psi _t} &= K\log ( {1 + \frac{{{\theta _{1}}}}{{\sigma _t^2}}} ) + {\theta _{2}}. \label{psi}
\end{align}
Note that, different from the learned denoising-based approximate message passing (LDAMP) network [36], where the authors replaced the denoiser module ${D_{{{\hat \sigma }^l}}}(  \cdot  )$ in the DAMP algorithm with the denoising convolutional neural network (DnCNN), we derive the shrinkage function $\eta ( { \cdot ; \cdot } )$ in detail, which plays the role of the nonlinear activation function in deep learning.
Moreover, instead of processing the element $r$ in the existing MDDL-based scheme \cite{LAMP_TSP, HHT_WC,HHT_LDAMP,AMP_WCL},
the developed MMV-LAMP network processes the row vector ${{\mathbf{r}}_j}$ for $1 \le j \le N$ by fully exploiting the structured sparsity of ${\mathbf{X}}$ from the {\it{a priori}} model. 
The derivation of the developed MMV-LAMP network is shown in Appendix.
Additionally, in order to avoid the performance degradation caused by the mismatch between the continuous angles and the discrete dictionary, we integrate the redundant dictionary into the CRN (i.e., the decoder) to improve the channel estimation performance.

\begin{figure*}[t]
	\centering
	\includegraphics[scale=0.35]{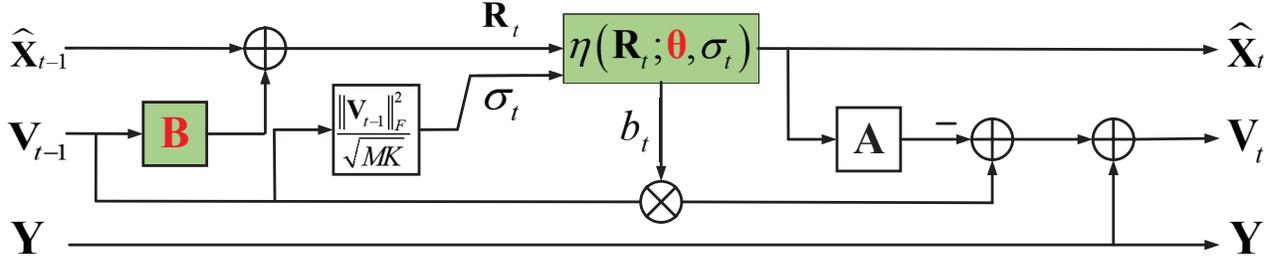}
	\caption{The $t$-th layer architecture of the developed MMV-LAMP network with the trainable parameters $\left\{ {{\mathbf{B}},{\bm{\theta }}} \right\}$.}
	\label{LAMP}
\end{figure*}

Fig. \ref{LAMP} illustrates the $t$-th layer architecture of the developed MMV-LAMP network.
In Fig. \ref{LAMP}, the inputs are ${{\mathbf{\widehat X}}_{t - 1}} \in {\mathbb{C}^{N \times K}}$, ${{\mathbf{V}}_{t - 1}} \in {\mathbb{C}^{M \times K}}$, and ${\mathbf{Y}} \in {\mathbb{C}^{M \times K}}$, where ${{\mathbf{\widehat X}}_{t - 1}}$ and ${{\mathbf{V}}_{t - 1}}$ are the outputs of the previous $( {t - 1} )$-th layer, and $\mathbf{Y}$ is the noisy measurement in (\ref{MMV_problem}).
Moreover, at the network training stage, we define ${\mathbf{B}} \in {\mathbb{C}^{N \times M}}$ and ${\bm{\theta }} = \left\{ {{\theta _1},{\theta _2}} \right\}$ as the trainable parameters of the MMV-LAMP network, which are identical for all $T$ layers.

\subsection{MMV-LAMP Network Based Uplink Channel Estimation}
\begin{figure*}[t]
	\centering
	\includegraphics[width=6.68in]{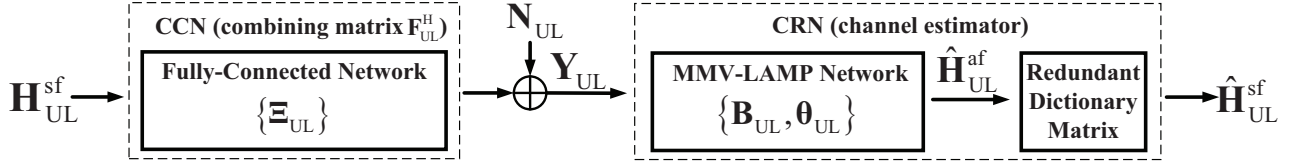}
	\caption{The block diagram of the proposed MDDL-based uplink channel estimation solution, which includes a CCN and an MMV-LAMP network based CRN.}
	\label{CE_LAMP}
\end{figure*}

The block diagram of the proposed MMV-LAMP network based uplink channel estimation scheme is depicted in Fig. \ref{CE_LAMP}, which contains the CCN and CRN.
The CCN corresponds to the combining matrix ${\mathbf{F}}_{{\text{UL}}}^{\text{H}}$ (encoder) in (\ref{uplink received signal K}), and the CRN corresponds to the channel estimator (decoder) at the BS.
Specifically, the input and output of the CCN are the uplink spatial-frequency domain channel matrix ${\mathbf{H}}_{{\text{UL}}}^{{\text{sf}}}$ and the noiseless pilot signals received at the BS.
Note that the parameters of the CCN $\left\{ {\bm{\Xi }}_{\text{UL}} \right\}$ corresponds to the phase values of the combining matrix ${\mathbf{F}}_{{\text{UL}}}^{\text{H}}$ in (\ref{uplink received signal K}).
Moreover, the CRN consists of $T$ layers and each has the same network structure and trainable parameters $\left\{ {{\mathbf{B}}_{\text{UL}},{\bm{\theta }}}_{\text{UL}} \right\}$ as shown in Fig. \ref{LAMP}, whose output is the estimated angle-frequency domain channel matrix $\widehat {\mathbf{H}}_{{\text{UL}}}^{{\text{af}}}$.
At the offline training stage, we jointly train the overall network parameters $\left\{ {\bm{\Xi }}_{\text{UL}},{\mathbf{B}}_{\text{UL}},{\bm{\theta }}_{\text{UL}} \right\}$ as an auto-encoder in an end-to-end approach.
Finally, the estimated spatial-frequency domain channel matrix ${\widehat {\mathbf{H}}_{{\text{UL}}}^{{\text{sf}}}}$ can be obtained by multiplying a devised redundant dictionary matrix.

\subsubsection{Fully-Connected CCN}
\begin{figure}[t]
	\centering
	\includegraphics[scale=0.3]{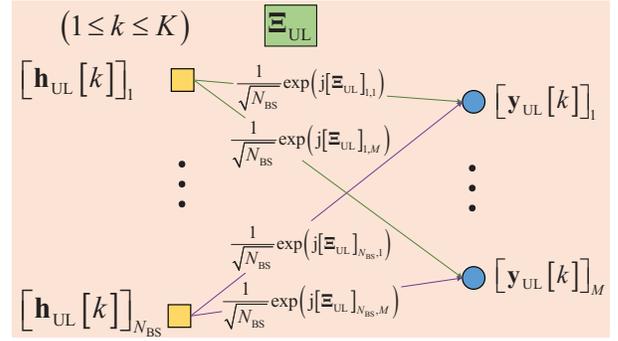}
	\caption{The proposed fully-connected CCN with the trainable parameters $\left\{ {\bm{\Xi}}_{\text{UL}} \right\}$, which correspond to the combining matrix ${\mathbf{F}}_{{\text{UL}}}^{\text{H}}$.}
	\label{dimension reduction network}
\end{figure}

Consider the formula ${{\mathbf{Y}}_{{\text{UL}}}} = {\mathbf{F}}_{{\text{UL}}}^{\text{H}}{\mathbf{H}}_{{\text{UL}}}^{{\text{sf}}} + { {\mathbf{N}}}_{\text{UL}}$ in (\ref{uplink received signal K}), in order to mimic the linear compressibility process of high-dimensional channels, the combining matrix ${\mathbf{F}}_{{\text{UL}}}$ can be well modeled as a CCN realized by a fully-connected layer without biases and a nonlinear activation function.
Note that the combining matrix ${\mathbf{F}}_{{\text{UL}}}$ is a complex-valued matrix and satisfies the constant modulus constraint due to the RF PSN adopted in the hybrid MIMO architecture, and the expression of the combining matrix ${\mathbf{F}}_{{\text{UL}}}$ is given by
\begin{align}\label{W}
	{{\mathbf{F}}_{{\text{UL}}}}{\text{ }} = \frac{1}{{\sqrt {{N_{{\text{BS}}}}} }}\exp( {{\text{j}}{{\bm{\Xi }}_{{\text{UL}}}}} ) 
	             = \frac{1}{{\sqrt {{N_{{\text{BS}}}}} }}\left[ {\cos ( {\mathbf{\Xi }}_{\text{UL}} ) + {\text{j}}\sin ( {\mathbf{\Xi }}_\text{UL} )} \right],
\end{align}
where ${\text{j}} = \sqrt { - 1}$ and ${\left[ {\mathbf{\Xi }}_{\text{UL}} \right]_{m,n}} \in \left[ {0,2\pi } \right)$.
As it is well known that complex-valued outputs are not well supported by most deep learning frameworks (e.g., Tensorflow, Pytorch), it would be difficult to directly train the complex-valued combining matrix ${\mathbf{F}}_{{\text{UL}}}$.
Hence, for the fully-connected CCN, we choose to train the real-valued phases of PSN $\left\{ {\bm{\Xi}}_{\text{UL}} \right\}$, in other words, we define the real-valued phases of PSN as the real-valued trainable parameters of the fully-connected CCN.
Moreover, the structure of the proposed fully-connected CCN is shown in Fig. \ref{dimension reduction network}, where the trainable parameter of the CCN is $\left\{ {\bm{\Xi}}_{\text{UL}} \right\}$ and the corresponding weight matrix of the CCN is $\exp ( {{\text{j}}{\bm{\Xi} _{{\text{UL}}}}} )/\sqrt {{N_{{\text{BS}}}}} $.
Hence, the parameters of the fully-connected layer are regarded as the phases of PSN and can be learned at the deep learning training stage.

\subsubsection{CRN Based on MMV-LAMP Network}
First, we detail the devised redundant dictionary matrix.
Specifically, massive MIMO channels are sparse in the angle-domain, and the accuracy of CRN depends heavily on their sparsity, which may be weakened by the power leakage \cite{WZW_TVT}.
Therefore, we design a redundant dictionary matrix ${\mathbf{D}}$ with a finer angular resolution to transform the spatial-frequency domain channel matrix ${{\mathbf{H}}^{{\text{sf}}}}$ into the angle-frequency domain channel matrix ${{\mathbf{H}}^{{\text{af}}}}$, which can be expressed as
\begin{equation}\label{angular domain channel}
{{\mathbf{H}}^{\text{sf}}} = {{\mathbf{D}}^{\text{H}}}{{\mathbf{H}}^{\text{af}}},
\end{equation}
where the redundant dictionary matrix ${\mathbf{D}} \in {\mathbb{C}^{G \times {N_{{\text{BS}}}}}}$ consists of $G$ column vectors ${\mathbf{a}}( {{\phi _g}} )$ for $1 \le g \le G$, i.e., ${\mathbf{D}} = {\left[ {{\mathbf{a}}( {{\phi _1}} ),{\mathbf{a}}( {{\phi _2}} ), \cdots ,{\mathbf{a}}( {{\phi _G}} )} \right]^{\text{T}}}$, with ${\mathbf{a}}( {{\phi _g}} )$ the corresponding array steering vector in (\ref{steeting vector}), where the sine function in the array steering vector is defined as $\sin ( {{\phi _g}} ) =  - 1 + 2( {g - 1})/G$ by quantifying the range of AoDs into $G$ grids for $g = 1,2, \cdots ,G$.
Therefore, estimating ${\mathbf{H}}_{{\text{UL}}}^{{\text{sf}}}$ in (\ref{uplink received signal K}) is equivalent to estimating ${\mathbf{H}}_{{\text{UL}}}^{{\text{af}}}$ represented in the angle-domain redundant dictionary $\mathbf{D}$, i.e.,
\begin{align}\label{LAMP_CE_problem}
	{{\mathbf{Y}}_{{\text{UL}}}} = {\mathbf{F}}_{{\text{UL}}}^{\text{H}}{{\mathbf{D}}^{\text{H}}}{\mathbf{H}}_{{\text{UL}}}^{{\text{af}}} + {{{\mathbf{ N}}}_{{\text{UL}}}} 
	             = {\mathbf{A}}_{\text{UL}}{\mathbf{H}}_{{\text{UL}}}^{{\text{af}}} + {{{\mathbf{ N}}}_{{\text{UL}}}},
\end{align}
where ${\mathbf{A}}_{\text{UL}} = {\mathbf{F}}_{{\text{UL}}}^{\text{H}}{{\mathbf{D}}^{\text{H}}}$ is the effective measurement matrix.
It's worth noting that the $k$-th column of ${\mathbf{H}}_{{\text{UL}}}^{{\text{af}}}$, i.e., ${\mathbf{H}}_{{\text{UL}}}^{{\text{af}}}( {:,k} )$ is a sparse column vector, meanwhile, $\big\{ {{\mathbf{H}}_{{\text{UL}}}^{{\text{af}}}( {:,k} )} \big\}_{k = 1}^K$ share the common sparsity \cite{GZ_TSP}.
Consequently, the sparse channel estimation problem can be formulated as an MMV sparse matrix recovery problem in CS. Noth that, given the received signals, the channel matrix ${\mathbf{H}}_{{\text{UL}}}^{{\text{af}}}$ can be estimated by solving the following optimization problem
\begin{align}\label{CS_optimization}
\mathop {\min }\limits_{{\mathbf{H}}_{{\text{UL}}}^{{\text{af}}}} &{\left( {\sum\limits_{k = 1}^K {\left\| {{\mathbf{H}}_{{\text{UL}}}^{{\text{af}}}( {:,k} )} \right\|_0^2} } \right)^{1/2}} \\
{\text{s}}{\text{.t}}{\text{.}} &{{\left\| {{{\mathbf{Y}}_{{\text{UL}}}} - {{\mathbf{A}}_{{\text{UL}}}}{\mathbf{H}}_{{\text{UL}}}^{{\text{af}}}} \right\|_F} \le \delta}, \nonumber \\
{\text{and}} {\ } {\big\{ {{\mathbf{H}}_{{\text{UL}}}^{{\text{af}}}( {:,k} )} \big\}_{k = 1}^K}{\ }&{\text{share the common sparse support set}},  \nonumber
\end{align}
where ${\left\| {{\mathbf{H}}_{{\text{UL}}}^{{\text{af}}}( {:,k} )} \right\|_0}$ is the number of non-zero elements of ${{\mathbf{H}}_{{\text{UL}}}^{{\text{af}}}( {:,k} )}$ and $\delta$ is the error tolerance parameter.
By replacing the ${l_0}$-norm with the ${l_1}$-norm, various CS algorithms can be utilized to solve the problem, such as the simultaneous orthogonal matching pursuit (SOMP) algorithm \cite{GZ_WC2}, the MMV-AMP algorithm \cite{KML_TSP}, and the proposed MMV-LAMP algorithm.
However, these greedy CS algorithms cannot achieve satisfactory channel estimation accuracy.

\begin{figure*}[t]
	\centering
	\includegraphics[scale=0.71]{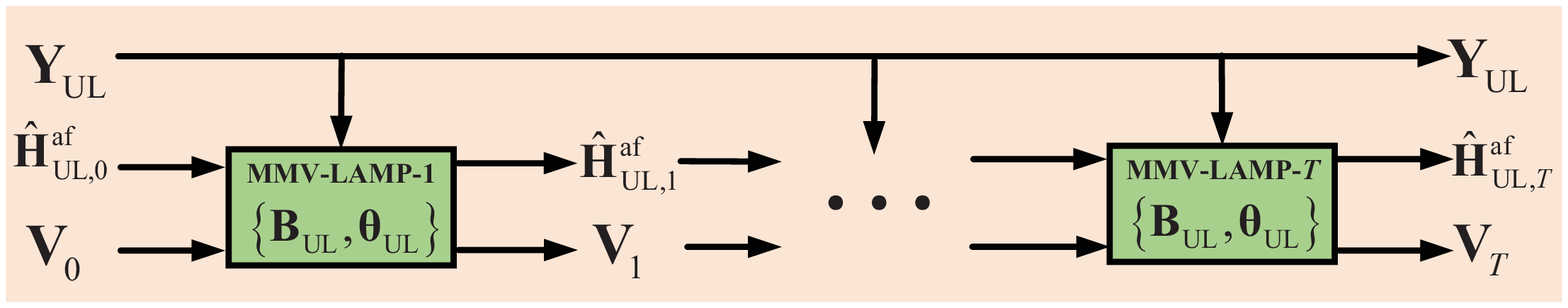}
	\caption{The proposed CRN based on MMV-LAMP network with the trainable parameters $\left\{ {{\mathbf{B}}_{\text{UL}},{\bm{\theta }}_{\text{UL}}} \right\}.$}
	\label{reconstruction network}
\end{figure*}

To efficiently solve the MMV CS problem in (\ref{CS_optimization}), we further develop an MMV-LAMP network with $T$ layers as illustrated in Fig. \ref{reconstruction network} and  summarized in Algorithm \ref{CE_LAMP_Algorithm}, which can reconstruct the high-dimensional angle-frequency domain channel matrix $\widehat {{\mathbf{H}}}_{{\text{UL}}}^{{\text{af}}}$ from the low-dimensional received signals ${{\mathbf{Y}}_{{\text{UL}}}}$.
Specifically, the input is the received signals ${{\mathbf{Y}}_{{\text{UL}}}}$ and the output of the $t$-th layer is the estimated angle-frequency domain channel matrix $\widehat {\mathbf{H}}_{{\text{UL}},t}^{{\text{af}}}$.
In addition, the initial values $\widehat {\mathbf{H}}_{{\text{UL}},0}^{{\text{af}}}$ and ${{\mathbf{V}}_0}$ are denoted as $\widehat {\mathbf{H}}_{{\text{UL}},0}^{{\text{af}}} = {\mathbf{0}}$ and ${{\mathbf{V}}_0} = {{\mathbf{Y}}_{{\text{UL}}}}$, respectively.

\begin{algorithm}[t]
	\caption{MMV-LAMP network based CRN}
	\label{CE_LAMP_Algorithm}
	\begin{algorithmic}[1]
		\REQUIRE The received signal ${\mathbf{Y}_{\text{UL}}}$, the measurement matrix ${\mathbf{A}}_{\text{UL}}$, the number of layers $T$.
		\ENSURE The output of the $T$-th layer MMV-LAMP network
		${\widehat {\mathbf{H}}_{{\text{UL}}}^{{\text{sf}}}} = {{\mathbf{D}}^{\text{H}}}{\widehat {\mathbf{H}}^{{\text{af}}}_{{\text{UL}},T}}$.
		\STATE Initialization: ${{\mathbf{V}}_0} = {\mathbf{Y}_{\text{UL}}}$, ${\widehat {\mathbf{H}}_{{\text{UL,}}0}^{{\text{af}}}} = {\mathbf{0}}$, ${\mathbf{B}}_{\text{UL}} = {{\mathbf{A}}_{\text{UL}}^{\text{H}}}$, ${\bm{\theta }}_{\text{UL}} = \left\{ {1,1} \right\}$.
		\FOR {$t = 1,2, \cdots ,T$}
			\STATE ${{\mathbf{R}}_t} = {\widehat {\mathbf{H}}_{{\text{UL,}}t-1}^{{\text{af}}}}+ {\mathbf{B}}_{\text{UL}}{{\mathbf{V}}_{t - 1}}$
			\STATE ${\sigma _t} = \frac{1}{{\sqrt {QK} }}{\left\| {{{\mathbf{V}}_{t - 1}}} \right\|_F}$
			\STATE ${\widehat {\mathbf{H}}_{{\text{UL,}}t}^{{\text{af}}}} = \eta ( {{{\mathbf{R}}_t};{\bm{\theta }}_{\text{UL}},{\sigma _t}} )$
			\STATE ${b_t} = \frac{1}{Q}\sum\limits_{j = 1}^G {\frac{{\partial {{\left[ {{{\eta}} ( {{{\mathbf{R}}_t};{{\bm{\theta}}}_{\text{UL}},{\sigma _t}} )} \right]}_j}}}{\partial \left[ {{{\mathbf{R}}_t}( {j,:} )} \right]}}$
			\STATE ${{\mathbf{V}}_t} = {\mathbf{Y}_{\text{UL}}} - {\mathbf{A}}_{\text{UL}}{\widehat {\mathbf{H}}_{{\text{UL,}}t}^{{\text{af}}}} + {b_t}{{\mathbf{V}}_{t - 1}}$
		\ENDFOR
	\end{algorithmic}
\end{algorithm}

As for the $t$-th layer, the trainable parameter $\left\{ {\mathbf{B}}_{\text{UL}} \right\}$ is denoted as
\begin{equation}\label{B}
	{\mathbf{B}}_{\text{UL}} = \operatorname{Re} \left\{ {\mathbf{B}}_{\text{UL}} \right\} + {\text{j}}\operatorname{Im} \left\{ {\mathbf{B}}_{\text{UL}} \right\},
\end{equation}
where $\operatorname{Re} \left\{ {\mathbf{B}}_{\text{UL}} \right\}$ and $\operatorname{Im} \left\{ {\mathbf{B}}_{\text{UL}} \right\}$ denote the real and imaginary parts of the trainable parameter ${\mathbf{B}}_{\text{UL}}$, respectively.
In other words, in order to achieve mathematical complex-valued processing in the MMV-LAMP network, we define two real-valued trainable parameters $\operatorname{Re} \left\{ {\mathbf{B}}_{\text{UL}} \right\}$ and $\operatorname{Im} \left\{ {\mathbf{B}}_{\text{UL}} \right\}$ to form the complex-valued trainable parameters $\left\{ {\mathbf{B}}_{\text{UL}} \right\}$.
Finally, the final estimated spatial-frequency domain channel matrix based on the output of the $T$-th layer is given by
\begin{equation}\label{estimated spatial-frequency domain channel}
	{\widehat {\mathbf{H}}_{{\text{UL}}}^{{\text{sf}}}} = {{\mathbf{D}}^{\text{H}}}\widehat {\mathbf{H}}_{{\text{UL}},T}^{{\text{af}}}.
\end{equation}

\subsubsection{Learning Strategy}
Inspired by the auto-encoder, we propose a novel layer-by-layer learning strategy to
jointly train the CCN (encoder) and CRN (decoder).
Specifically, at the offline training stage, we first generate the training data set $\big\{ {{\mathbf{H}}_{{\text{UL}}}^{{\text{sf,}}n}} \big\}_{n = 1}^{{N_{{\text{train}}}}}$ according to (\ref{frequency channel}),
where ${{N_{{\text{train}}}}}$ is the number of channel samples in the training set,
and ${{{\mathbf{H}}^{{\text{sf}},n}_{\text{UL}}}}$ is not only the input of the CCN, but also the corresponding target output.
In order to jointly optimize the trainable parameters of the CCN and CRN, i.e., $\left\{ {\bm{\Xi }}_{\text{UL}},{\mathbf{B}}_{\text{UL}},{\bm{\theta }}_{\text{UL}} \right\}$,
we define the normalized mean square error (NMSE) between the target value ${{{\mathbf{H}}^{{\text{sf}},n}_{\text{UL}}}}$ and the estimated spatial-frequency channel matrix of the $t$-th layer MMV-LAMP network $\widehat {\mathbf{H}}_{{\text{UL,}}t}^{{\text{sf}},n} = {{\mathbf{D}}^{\text{H}}}\widehat {\mathbf{H}}_{{\text{UL,}}t}^{{\text{af}},n}$ as the loss function of the $t$-th layer $\footnote{The ultimate goal of the proposed MDDL-based channel estimation scheme is to obtain better NMSE performance, hence, we choose the NMSE rather than the MSE as the loss function.}$, i.e., 
\begin{align}\label{NMSE_t}
	{L_{\text{UL},t}} &\Big( {\big\{ {{{\bm{\Xi }}_{{\text{UL}}}},{{\mathbf{B}}_{{\text{UL}}}},{{\bm{\theta }}_{{\text{UL}}}}} \big\}_t} \Big) \nonumber \\
	&= \sum\limits_{n = 1}^N {\frac{{\left\| {\widehat {\mathbf{H}}_{{\text{UL}},t}^{{\text{sf}},n} - {\mathbf{H}}_{{\text{UL}}}^{{\text{sf}},n}} \right\|_F^2}}{{\left\| {{\mathbf{H}}_{{\text{UL}}}^{{\text{sf}},n}} \right\|_F^2}}} \nonumber \\
	&= \sum\limits_{n = 1}^N {\frac{{\left\| {{{\mathbf{D}}^{\text{H}}}}{{f_t}( {{\mathbf{H}}_{{\text{UL}}}^{{\text{sf}},n},{\big\{ {{{\bm{\Xi }}_{{\text{UL}}}},{{\mathbf{B}}_{{\text{UL}}}},{{\bm{\theta }}_{{\text{UL}}}}} \big\}_t}} ) - {\mathbf{H}}_{{\text{UL}}}^{{\text{sf}},n}} \right\|_F^2}}{{\left\| {{\mathbf{H}}_{{\text{UL}}}^{{\text{sf}},n}} \right\|_F^2}}},
\end{align}
where $N$ is the number of data samples in each batch of the training set, ${\mathbf{H}}_{{\text{UL}}}^{{\text{sf}},n}$ is the $n$-th uplink spatial-frequency domain channel sample,
and ${\widehat {\mathbf{H}}_{{\text{UL}},t}}^{{\text{af}},n} = {{f_t}( {{\mathbf{H}}_{{\text{UL}}}^{{\text{sf}},n},{\left\{ {{{\mathbf{\Xi }}_{{\text{UL}}}},{{\mathbf{B}}_{{\text{UL}}}},{{\mathbf{\theta }}_{{\text{UL}}}}} \right\}_t}} )}$ is the output of the $t$-th layer MMV-LAMP network.
Note that ${f_t}( { \cdot , \cdot } )$ indicates the proposed uplink channel estimation solution including the CCN and CRN, where the CRN is iterated $t$ times in the $t$-th layer, i.e., ${f_t}( { \cdot , \cdot } ) = f^{{\text{CRN}}}( { \cdots f^{{\text{CRN}}}( {f^{{\text{CCN}}}( {{\mathbf{H}}_{{\text{UL}}}^{{\text{sf,}}n},{{\bm{\Xi }}_{{\text{UL}}}}} ),{{\mathbf{B}}_{{\text{UL}}}},{{\bm{\theta }}_{{\text{UL}}}}} ),{{\mathbf{B}}_{{\text{UL}}}},{{\bm{\theta }}_{{\text{UL}}}}} )$.

The proposed novel layer-by-layer learning strategy is summarized in Algorithm 2, and the Adam algorithm with the learning rate 0.001 is adopted \cite{Adam}. 
Specifically, for the 1-st layer, the input data is the target data ${{{\mathbf{H}}^{{\text{sf}}}_{\text{UL}}}}$ and the output is $\widehat {\mathbf{H}}_{{\text{UL,}}1}^{{\text{sf}}} = {{\mathbf{D}}^{\text{H}}}{f^{{\text{CRN}}}}( {{f^{{\text{CCN}}}}( {{\mathbf{H}}_{{\text{UL}}}^{{\text{sf}}},{{\mathbf{\Xi }}_{{\text{UL}}}}} ),{{\mathbf{B}}_{{\text{UL}}}},{{\bm{\theta }}_{{\text{UL}}}}} )$, where ${f^{{\text{CCN}}}}( { \cdot , \cdot } )$ and ${f^{{\text{CRN}}}}( { \cdot , \cdot , \cdot } )$ denote the proposed CCN and CRN structure, respectively.
Therefore, we aim to optimize the $1$-st layer's trainable parameters ${\left\{ {{{\bm{\Xi }}_{{\text{UL}}}},{{\mathbf{B}}_{{\text{UL}}}},{{\bm{\theta }}_{{\text{UL}}}}} \right\}_1}$ by minimizing the loss function of the $1$-st layer ${L_{\text{UL},1}} \big( {\left\{ {{{\bm{\Xi }}_{{\text{UL}}}},{{\mathbf{B}}_{{\text{UL}}}},{{\bm{\theta }}_{{\text{UL}}}}} \right\}_1} \big)$.
For the 2-nd layer, the input data is ${{{\mathbf{H}}^{{\text{sf}}}_{\text{UL}}}}$, but the output is $\widehat {\mathbf{H}}_{{\text{UL,2}}}^{{\text{sf}}} = {{\mathbf{D}}^{\text{H}}}{f^{{\text{CRN}}}}( {{f^{{\text{CRN}}}}( {{f^{{\text{CCN}}}}( {{\mathbf{H}}_{{\text{UL}}}^{{\text{sf}}},{{\bm{\Xi }}_{{\text{UL}}}}} ),{{\mathbf{B}}_{{\text{UL}}}},{{\bm{\theta }}_{{\text{UL}}}}} ),{{\mathbf{B}}_{{\text{UL}}}},{{\bm{\theta }}_{{\text{UL}}}}} )$.
And the trainable parameters and the loss function of the $2$-nd layer are ${\left\{ {{{\bm{\Xi }}_{{\text{UL}}}},{{\mathbf{B}}_{{\text{UL}}}},{{\bm{\theta }}_{{\text{UL}}}}} \right\}_2}$ and ${L_{\text{UL},2}} \big( {\left\{ {{{\bm{\Xi }}_{{\text{UL}}}},{{\mathbf{B}}_{{\text{UL}}}},{{\bm{\theta }}_{{\text{UL}}}}} \right\}_2} \big)$, respectively.
Note that, the initial values of the $2$-nd layer's trainable parameters ${\left\{ {{{\bm{\Xi }}_{{\text{UL}}}},{{\mathbf{B}}_{{\text{UL}}}},{{\bm{\theta }}_{{\text{UL}}}}} \right\}_2}$ are the obtained parameters ${\left\{ {{{\bm{\Xi }}_{{\text{UL}}}},{{\mathbf{B}}_{{\text{UL}}}},{{\bm{\theta }}_{{\text{UL}}}}} \right\}_1}$ from the $1$-st layer's training.
Similarly, for $T$-th layer, the input data is still the target data ${{{\mathbf{H}}^{{\text{sf}}}_{\text{UL}}}}$, and the output is $\widehat {\mathbf{H}}_{{\text{UL,}}T}^{{\text{sf}}}$, where $\widehat {\mathbf{H}}_{{\text{UL,}}T}^{{\text{sf}}} = {{\mathbf{D}}^{\text{H}}}{f^{{\text{CRN}}}}( { \cdots {f^{{\text{CRN}}}}( {{f^{{\text{CCN}}}}( {{\mathbf{H}}_{{\text{UL}}}^{{\text{sf}}},{{\bm{\Xi }}_{{\text{UL}}}}} ),{{\mathbf{B}}_{{\text{UL}}}},{{\bm{\theta }}_{{\text{UL}}}}} ), {{\mathbf{B}}_{{\text{UL}}}},{{\bm{\theta }}_{{\text{UL}}}}} )$.
Also, the trainable parameters and the loss function of the $T$-th layer are ${\left\{ {{{\bm{\Xi }}_{{\text{UL}}}},{{\mathbf{B}}_{{\text{UL}}}},{{\bm{\theta }}_{{\text{UL}}}}} \right\}_T}$ and ${L_{\text{UL},T}} \big( {\left\{ {{{\bm{\Xi }}_{{\text{UL}}}},{{\mathbf{B}}_{{\text{UL}}}},{{\bm{\theta }}_{{\text{UL}}}}} \right\}_T} \big)$, respectively.
Note that, the initial values of the $T$-th layer's trainable parameters are the obtained parameters ${\left\{ {{{\bm{\Xi }}_{{\text{UL}}}},{{\mathbf{B}}_{{\text{UL}}}},{{\bm{\theta }}_{{\text{UL}}}}} \right\}_{T-1}}$ from the $(T-1)$-th layer's training.
In other words, the proposed layer-by-layer training strategy combines both the conventional layer-by-layer and all-layer training strategy.
Consider the $t$-th layer training $( {1 \le t \le T} )$, we adopt the all-layer training strategy, i.e., the trainable parameters ${\left\{ {{{\bm{\Xi }}_{{\text{UL}}}},{{\mathbf{B}}_{{\text{UL}}}},{{\bm{\theta }}_{{\text{UL}}}}} \right\}_{t}}$ are jointly optimized; while from the perspective of $T$ times training with the increasing layer number $t$, it is a modified kind of layer-by-layer training strategy.
After the trainable parameters $\left\{ {{\bm{\Xi }}_{\text{UL}},{\mathbf{B}}_{\text{UL}},{\bm{\theta }}_{\text{UL}}} \right\}_{T}$ of the $T$-th layer are optimized,
we can obtain the complex-valued PSN and the MMV-LAMP network simultaneously, which can be adopted to design the combining matrix ${\mathbf{F}}_{{\text{UL}}}^{\text{H}}$ and the channel estimator at the online channel estimation stage.
\begin{figure*}[t]
	\centering
	\includegraphics[width=483pt]{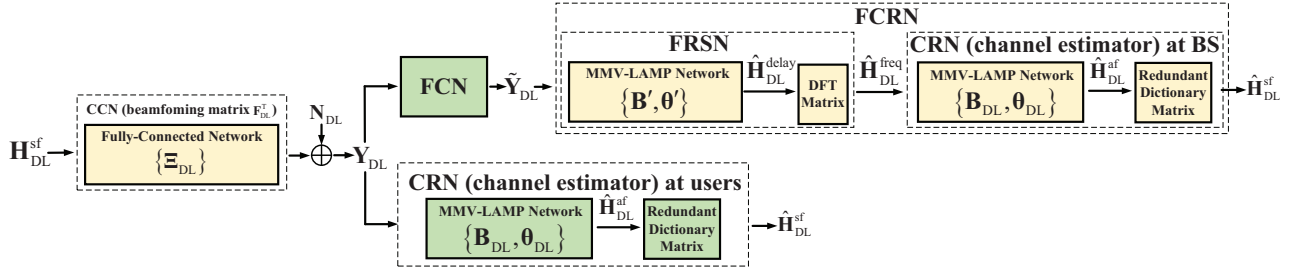}
	\caption{The block diagram of the proposed MDDL-based downlink channel estimation and feedback solution, where the green and yellow block diagrams represent that the modules are processed at the users and the BS, respectively.}
	\label{feedback_overall}
\end{figure*}

\begin{algorithm}[t]
	\caption{Learning strategy to jointly train CCN's parameters ${\left\{ {{{\bm{\Xi }}_{{\text{UL}}}}} \right\}}$ and CRN's parameters ${\left\{ {{{\mathbf{B}}_{{\text{UL}}}},{{\bm{\theta }}_{{\text{UL}}}}} \right\}}$}
	\label{learning parameters}
	\begin{algorithmic}[1]
		\STATE Initialization: ${\mathbf{B}}_{\text{UL}} = {{\mathbf{A}}_{\text{UL}}^{\text{H}}}$, ${\bm{\theta }}_{\text{UL}} = \left\{ {1,1} \right\}$, ${\left[ {\mathbf{\Xi }}_{\text{UL}} \right]_{i,j}} \in \left[ {0,2\pi } \right)$, ${\left\{ {{\bm{\Xi }}_{\text{UL}},{\mathbf{B}}_{\text{UL}},{\bm{\theta }}_{\text{UL}}} \right\}_0}=\left\{ {{\bm{\Xi }}_{\text{UL}},{\mathbf{B}}_{\text{UL}},{\bm{\theta }}_{\text{UL}}} \right\}$.
		\FOR {$t = 1,2, \cdots ,T$}
			\STATE Initialize ${\left\{ {{\bm{\Xi }}_{\text{UL}},{\mathbf{B}}_{\text{UL}},{\bm{\theta }}_{\text{UL}}} \right\}_t}$ as $\left\{ {{\bm{\Xi }}_{\text{UL}},{\mathbf{B}}_{\text{UL}},{\bm{\theta }}_{\text{UL}}} \right\}_{t-1}$
			\STATE Learn ${\left\{ {{\bm{\Xi }}_{\text{UL}},{\mathbf{B}}_{\text{UL}},{\bm{\theta }}_{\text{UL}}} \right\}_t}$ to minimize the loss function of $t$-th layer ${L_{\text{UL},t}}( {\left\{ {{\bm{\Xi }}_{\text{UL}},{\mathbf{B}}_{\text{UL}},{\bm{\theta }}_{\text{UL}}} \right\}_t} )$
		\ENDFOR
		\ENSURE $\left\{ {{\bm{\Xi }}_{\text{UL}},{\mathbf{B}}_{\text{UL}},{\bm{\theta }}_{\text{UL}}} \right\} = {\left\{ {{\bm{\Xi }}_{\text{UL}},{\mathbf{B}}_{\text{UL}},{\bm{\theta }}_{\text{UL}}} \right\}_T}.$
	\end{algorithmic}
\end{algorithm}

\section{MDDL-Based FDD Downlink Channel Estimation and Feedback}
In this section, we first extend the proposed MDDL-based TDD uplink channel estimation scheme to the FDD downlink channel estimation.
Moreover, since the uplink/downlink channel reciprocity does not hold in FDD systems,
we further propose an MMV-LAMP network based channel feedback solution, whereby the channels' delay-domain sparsity is exploited for reducing the feedback overhead.
As shown in Fig. \ref{feedback_overall}, the block diagram of the proposed downlink channel estimation and feedback solution contains a CCN at the BS, a feedback compression network (FCN) at the users, and a FCRN at the BS, where the FCRN further consists of a FRSN and a CRN (this CRN is the same as that in Fig. \ref{CE_LAMP}).

\subsection{MMV-LAMP Network Based Downlink Channel Estimation}
Similar to Section III, estimating the ${\mathbf{H}}_{{\text{DL}}}^{{\text{sf}}}$ in (\ref{downlink received signal K}) is equivalent to estimating ${\mathbf{H}}_{{\text{DL}}}^{{\text{af}}}$ represented in the angle-domain redundant dictionary ${{\mathbf{D}}}$, i.e., 
\begin{align}\label{downlink LAMP_CE_problem}
	{\mathbf{Y}_{\text{DL}}} = {{\mathbf{F}}_{\text{DL}}^{\text{T}}}{{\mathbf{D}}^{\text{H}}}{{\mathbf{H}}^{{\text{af}}}_{\text{DL}}} + {{\mathbf{N}}}_{\text{DL}} 
	             = {\mathbf{A}}_{\text{DL}}{{\mathbf{H}}^{{\text{af}}}_{\text{DL}}} + { {\mathbf{N}}}_{\text{DL}},
\end{align}
where ${\mathbf{A}}_{\text{DL}} = {{\mathbf{F}}_{\text{DL}}^{\text{T}}}{{\mathbf{D}}^{\text{H}}} \in {\mathbb{C}^{Q \times G}}$ is the measurement matrix in CS.
We can observe that (\ref{LAMP_CE_problem}) and  (\ref{downlink LAMP_CE_problem}) share a similar expression, hence, the proposed MMV-LAMP network for uplink channel estimation scheme at the BS can be used for the downlink channel estimation at the users.

Specifically, at the downlink channel estimation stage, the input of the CCN is the downlink spatial-frequency domain channel matrix ${{\mathbf{H}}^{{\text{sf}}}_{\text{DL}}}$ and the output is the received signals ${\mathbf{Y}_{\text{DL}}}$ at the user.
Similar to Section III, the trainable parameters of the CCN can be regarded as the real-valued phases of PSN, and the corresponding expression of the complex-valued beamforming matrix ${\mathbf{F}}_{\text{DL}}$ is given by
\begin{align}\label{F}
	{\mathbf{F}}_{\text{DL}} = \frac{1}{{\sqrt {{N_{{\text{BS}}}}} }}{{\mathbf{e}}^{{\text{j}}\left[ {{{\bm{\Xi }}_{{\text{DL}}}}} \right]}} 
	             = \frac{1}{{\sqrt {{N_{{\text{BS}}}}} }}\left[ {\cos ( {{{\bm{\Xi }}_{{\text{DL}}}}} ) + {\text{j}}\sin ( {{{\bm{\Xi }}_{{\text{DL}}}}} )} \right].
\end{align}
On the other hand, for the CRN at the users, we replace the inputs ${\mathbf{Y}}_{\text{UL}}$ and ${\mathbf{A}}_{\text{UL}}$ with ${\mathbf{Y}}_{\text{DL}}$ and ${\mathbf{A}}_{\text{DL}}$, respectively, and the other terms are the same as in the Algorithm \ref{CE_LAMP_Algorithm}.
As for the learning strategy, we train the trainable parameters $\left\{ {{\bm{\Xi }}_{\text{DL}},{\mathbf{B}}_{\text{DL}},{\bm{\theta }}_{\text{DL}}} \right\}$ in the downlink channel estimation based on Algorithm \ref{learning parameters} by replacing the inputs ${\mathbf{A}}_{\text{UL}}$ and $\left\{ {{\bm{\Xi }}_{\text{UL}},{\mathbf{B}}_{\text{UL}},{\bm{\theta }}_{\text{UL}}} \right\}$ with ${\mathbf{A}}_{\text{DL}}$ and $\left\{ {{\bm{\Xi }}_{\text{DL}},{\mathbf{B}}_{\text{DL}},{\bm{\theta }}_{\text{DL}}} \right\}$, respectively.
\begin{figure*}[t]
	\centering
	\includegraphics[scale=0.65]{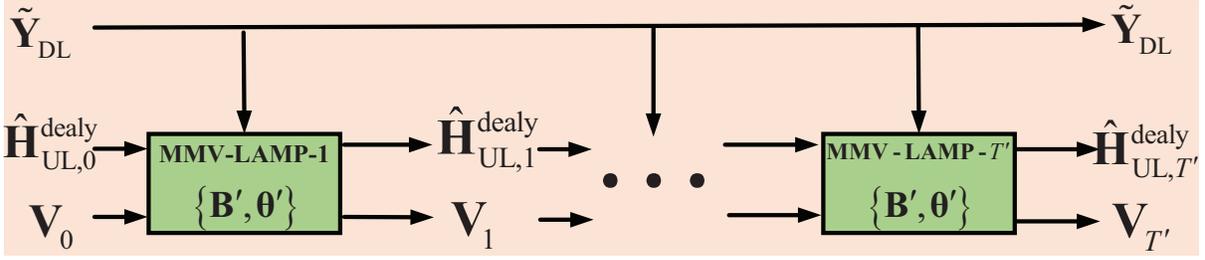}
	\caption{The proposed FRSN based on MMV-LAMP network with the trainable parameters $\left\{ {{\mathbf{B}'},{\bm{\theta }}'} \right\}$.}
	\label{FB_LAMP}
\end{figure*}

\subsection{MMV-LAMP Network Based Channel Feedback}
Given the received signal ${\mathbf{Y}}_{{\text{DL}}}$ in (\ref{downlink received signal K}) at the users, the received signals can be rewritten as
\begin{align}\label{FB_received_signal_Q}
	{\mathbf{Y}}_{{\text{DL}}}^{\text{T}} = {\mathbf{ H}}_{{\text{DL}}}^{{\text{fs}}}{{\mathbf{F}}_{{\text{DL}}}} + {\mathbf{N}}_{{\text{DL}}}^{\text{T}} 
	             =  {\mathbf{H}} _{{\text{DL}}}^{{\text{freq}}} + {\mathbf{N}}_{{\text{DL}}}^{\text{T}},
\end{align}
where ${\mathbf{ H}}_{{\text{DL}}}^{{\text{fs}}} = {( {{\mathbf{H}}_{{\text{DL}}}^{{\text{sf}}}} )^{\text{T}}} \in {\mathbb{C}^{K \times {N_{{\text{BS}}}}}}$ denotes the frequency-spatial domain channel matrix and ${\mathbf{H}}_{{\text{DL}}}^{{\text{freq}}} = {\mathbf{H}}_{{\text{DL}}}^{{\text{fs}}}{{\mathbf{F}}_{{\text{DL}}}} \in {\mathbb{C}^{K \times Q}}$ is a frequency-domain channel matrix after spatial-domain compression.

In order to accurately acquire the CSI at the BS with reduced feedback overhead, we propose an FCN and an FCRN based on an MMV-LAMP network with two steps.
In the first step, by exploiting the channels' delay-domain sparsity, we compress the received pilots at the users by only feeding back the received pilot signals on part of $K$ subcarriers.
In the second step, the compressed feedback signals received at the BS are regarded as the input of the FRSN to reconstruct the frequency-domain channel matrix $\widehat { {\mathbf{H}}}_{{\text{DL}}}^{{\text{freq}}}$,
which is then input to the CRN (can be well trained at the uplink channel estimation stage), so that the spatial-frequency domain channel matrix ${{\mathbf{H}}_{\text{DL}}^{{\text{sf}}}}$ is finally reconstructed at the BS.
\begin{algorithm}[t]
	\caption{MMV-LAMP network based FRSN}
	\label{FB_LAMP_Algorithm}
	\begin{algorithmic}[1]
		\REQUIRE The feedback signal ${\widetilde {\mathbf{Y}}}_{\text{DL}}$, the measurement matrix  \\ \qquad ${\widetilde {\mathbf{U}}}$, the number of layers $T'$.
		\ENSURE The output of the $T'$-th layer MMV-LAMP network \\ \qquad
		${\widehat { {\mathbf{H}}}_{{\text{DL}}}^{\text{freq}}} = {\mathbf{U}}\widehat { {\mathbf{H}}}_{{\text{DL}},T'}^{{\text{delay}}}.$
		\STATE Initialization: ${{\mathbf{V}}_0} = {\widetilde {\mathbf{Y}}}_{\text{DL}}$, ${\widehat { {\mathbf{H}}}_{{\text{DL,}}{0}}^{\text{delay}}} = {\mathbf{0}}$, ${\mathbf{B}'} = {\widetilde {\mathbf{U}}^{\text{H}}}$, ${\bm{\theta }'} = \left\{ {1,1} \right\}$.
		\FOR {$t = 1,2, \cdots ,T'$}
			\STATE ${{\mathbf{R}}_t} = {\widehat { {\mathbf{H}}}_{{\text{DL,}}{t-1}}^{\text{delay}}} + {\mathbf{B}'}{{\mathbf{V}}_{t - 1}}$
			\STATE ${\sigma _t} = \frac{1}{{\sqrt {K_cQ} }}{\left\| {{{\mathbf{V}}_{t - 1}}} \right\|_F}$
			\STATE ${\widehat { {\mathbf{H}}}_{{\text{DL,}}{t}}^{\text{delay}}} = \eta ( {{{\mathbf{R}}_t};{\bm{\theta }'},{\sigma _t}} )$
			\STATE ${b_t} = \frac{1}{K_c}\sum\limits_{j = 1}^K {\frac{{\partial {{\left[ {{{\eta}} ( {{{\mathbf{R}}_t};{{\bm{\theta}'}},{\sigma _t}} )} \right]}_j}}}{{\partial \left[ {{{\mathbf{R}}_t}( {j,:} )} \right]}}}$
			\STATE ${{\mathbf{V}}_t} = {\widetilde {\mathbf{Y}}}_{\text{DL}} - {\widetilde {\mathbf{U}}}{\widehat { {\mathbf{H}}}_{{\text{DL,}}{t}}^{\text{delay}}} + {b_t}{{\mathbf{V}}_{t - 1}}$
		\ENDFOR
	\end{algorithmic}
\end{algorithm}
\begin{algorithm}[t]
	\caption{Learning strategy to train FRSN's parameters $\left\{ {\mathbf{B}'},{\bm{\theta }'} \right\}$}
	\label{FB learning parameters}
	\begin{algorithmic}[1]
		\STATE Initialization: ${\mathbf{B}'} = {\widetilde{ {\mathbf{U}} }^{\text{H}}}$, ${\bm{\theta }'} = \left\{ {1,1} \right\}$, ${\left\{ {\mathbf{B}'},{\bm{\theta }}' \right\}_0} = \left\{ {\mathbf{B}'},{\bm{\theta }'} \right\}$.
		\FOR {$t = 1,2, \cdots ,T'$}
			\STATE Initialize ${\left\{ {\mathbf{B}'},{\bm{\theta }'} \right\}_t}$ as ${\left\{ {\mathbf{B}'},{\bm{\theta }'} \right\}_{t - 1}}$
			\STATE Learn ${\left\{ {\mathbf{B}'},{\bm{\theta }'} \right\}_t}$ to minimize ${L_t'}\big( {{{\left\{ {\mathbf{B}'},{\bm{\theta }'} \right\}}_t}} \big)$
		\ENDFOR
		\ENSURE ${\left\{ {\mathbf{B}'},{\bm{\theta }'} \right\}} = {\left\{ {\mathbf{B}'},{\bm{\theta }'} \right\}_{T'}}.$
	\end{algorithmic}
\end{algorithm}

\subsubsection{Feedback Compression Network at Users}
To reduce the channel feedback overhead at the users, we compress the dimensionality of the received pilot signals by exploiting the channels' delay-domain sparsity.
Specifically, we transform the frequency-spatial domain channel matrix ${ {\mathbf{H}}_{{\text{DL}}}^{{\text{fs}}}}$ into the delay-spatial domain channel matrix ${\mathbf{H}}_{{\text{DL}}}^{{\text{ds}}}$ via a DFT matrix ${\mathbf{U}} \in {\mathbb{C}^{K \times K}}$, and (\ref{FB_received_signal_Q}) can be expressed as
\begin{align}\label{FB_delay_received_signal}
	{\mathbf{Y}}_{{\text{DL}}}^{\text{T}} = {\mathbf{U}}{\mathbf{H}}_{{\text{DL}}}^{{\text{ds}}}{\mathbf{F}}_{\text{DL}} + {\mathbf{N}}_{{\text{DL}}}^{\text{T}} 
	= {\mathbf{U}}{ {\mathbf{H}}_{{\text{DL}}}^{{\text{delay}}}}  + {\mathbf{N}}_{{\text{DL}}}^{\text{T}},
\end{align}
where ${ {\mathbf{H}}_{{\text{DL}}}^{{\text{delay}}}} = {\mathbf{H}}_{{\text{DL}}}^{{\text{ds}}}{\mathbf{F}}_{\text{DL}} \in {\mathbb{C}^{K \times Q}}$ is a delay-domain channel matrix after spatial-domain compression.
The compressed pilot signals fed back to the BS can be expressed as
\begin{equation}\label{FB_signal}
	{\widetilde {\mathbf{Y}}_{{\text{DL}}}} = {\left. {{\mathbf{Y}}_{{\text{DL}}}^{\text{T}}} \right|_{\mathbf{\Omega }}} = {\mathbf{\widetilde U}}{ {\mathbf{H}}_{{\text{DL}}}^{{\text{delay}}}} + {\widetilde {\mathbf{N}}_{{\text{DL}}}},
\end{equation}
where 
${\mathbf{\widetilde U}} = {\left. {\mathbf{U}} \right|_{\mathbf{\Omega }}} \in {\mathbb{C}^{{K_c} \times K}}$ is a partial DFT matrix, ${\widetilde {\mathbf{N}}_{{\text{DL}}}} = {\left. {{\mathbf{N}}_{{\text{DL}}}^{\text{T}}} \right|_{\mathbf{\Omega }}}$, 
and the $i$-th element of the set ${\left\{ {\mathbf{\Omega }} \right\}_i}$ for $1 \le i \le {K_c}$ is randomly selected without repeating \cite{GZ_EL}.

\subsubsection{Feedback Based Channel Reconstruction Network at BS}
The FCRN consists of an FRSN and a CRN.
Different from the MMV-LAMP network with $T$ layers used in CRN,
we just need to train the FRSN based on an MMV-LAMP network with $T'$ layers to obtain the channel matrix $\widehat { {\mathbf{H}}}_{{\text{DL}}}^{{\text{freq}}}$,
which is illustrated in Fig. \ref{FB_LAMP} and summarized in Algorithm \ref{FB_LAMP_Algorithm}.
Specifically, the input is the feedback pilot signals ${\widetilde {\mathbf{Y}}_{{\text{DL}}}}$,
and the initial values $\widehat { {\mathbf{H}}}_{{\text{DL,}}0}^{{\text{delay}}}$ and ${{\mathbf{V}}_0}$ of Algorithm \ref{FB_LAMP_Algorithm} are denoted as $\widehat { {\mathbf{H}}}_{{\text{DL,}}0}^{{\text{delay}}} = {\mathbf{0}}$ and ${{\mathbf{V}}_0} = {\widetilde {\mathbf{Y}}_{{\text{DL}}}}$, respectively.
In FRSN, the measurement matrix is the partial DFT matrix ${\mathbf{\widetilde U}}$
and the trainable parameters are $\left\{ {{\mathbf{B}'},{\bm{\theta }'}} \right\}$.
After the BS receives the compressed feedback signals ${\widetilde {\mathbf{Y}}_{{\text{DL}}}}$ in (\ref{FB_signal}),
we first exploit the proposed FRSN to reconstruct the channel matrix $\widehat { {\mathbf{H}}}_{{\text{DL}}}^{{\text{freq}}} = {\mathbf{U}}\widehat { {\mathbf{H}}}_{{\text{DL}},T'}^{{\text{delay}}}$,
which is then passed to the CRN to reconstruct $\widehat { {\mathbf{H}}}_{{\text{DL}}}^{{\text{sf}}}$ based on MMV-LAMP network.

Moreover, the training strategy of the trainable parameters $\left\{ {{\mathbf{B}'},{\bm{\theta }'}} \right\}$ of the FRSN is summarized in Algorithm \ref{FB learning parameters}, where the corresponding loss function of $t$-th layer ($1 \le t \le T'$) is given by
\begin{align}\label{FB_loss_function}
{ L_t'}\Big( {\big\{ {{\mathbf{B}}',{\bm{\theta}}'} \big\}_t} \Big) &= \sum\limits_{n = 1}^{N'} {\frac{{\left\| {\widehat { {\mathbf{H}}}_{{\text{DL}},t}^{{\text{freq}},n} -  {\mathbf{H}}_{{\text{DL}}}^{{\text{freq}},n}} \right\|_F^2}}{{\left\| { {\mathbf{H}}_{{\text{DL}}}^{{\text{freq}},n}} \right\|_F^2}}}  \nonumber \\
	&= \sum\limits_{n = 1}^{ N'} {\frac{{\left\| {{\mathbf{U}}}{{{ f}_t'}( { {\widetilde {\mathbf{Y}}_{{\text{DL}}}^n},{\big\{ {{\mathbf{B}}',{\bm{\theta}}'} \big\}_t}} ) -  {\mathbf{H}}_{{\text{DL}}}^{{\text{freq}},n}} \right\|_F^2}}{{\left\| { {\mathbf{H}}_{{\text{DL}}}^{{\text{freq}},n}} \right\|_F^2}}},
\end{align}
where $N'$ is the number of data samples in each batch of the training set, and $f'{( { \cdot , \cdot } )_t}$ denotes the MMV-LAMP network based FRSN.
Specifically, we first generate the training data set $\big\{ {{\mathbf{H}}_{{\text{DL}}}^{{\text{freq,}}n}} \big\}_{n = 1}^{N'}$ according to (\ref{FB_received_signal_Q}).
Then, for the $t$-th layer, we aim to optimize the trainable parameters $\left\{ {{\mathbf{B}'},{\bm{\theta }'}} \right\}_{t}$ by minimizing the loss function of the $t$-th layer ${L_t'}( {\big\{ {{\mathbf{B'}},{\bm{\theta '}}} \big\}_t} )$ for $1 \le t \le T'$.
Finally, after the trainable parameters $\left\{ {{\mathbf{B}'},{\bm{\theta }'}} \right\}_{T'}$ of the $T'$-th layer are optimized,
we can obtain $\widehat { {\mathbf{H}}}_{{\text{DL}}}^{{\text{delay}}}$ and $\widehat { {\mathbf{H}}}_{{\text{DL}}}^{{\text{freq}}} = {\mathbf{U}}\widehat { {\mathbf{H}}}_{{\text{DL}}}^{{\text{delay}}}$.
Also, we can find that ${ {\mathbf{H}}}_{{\text{DL}}}^{{\text{freq}}}$ in (\ref{FB_received_signal_Q}) can also be expressed as
\begin{align}\label{FB_2CE}
{( { {\mathbf{H}}_{{\text{DL}}}^{{\text{freq}}}} )^{\text{T}}} = {{\mathbf{F}}_{\text{DL}}^{\text{T}}}{{\mathbf{H}}_{\text{DL}}^{\text{sf}}} = {{\mathbf{F}}_{\text{DL}}^{\text{T}}}{{\mathbf{D}}^{\text{H}}}{{\mathbf{H}}_{\text{DL}}^{\text{af}}} 
	= {\mathbf{A}}_{\text{DL}}{{\mathbf{H}}_{\text{DL}}^{{\text{af}}}},
\end{align}
which indicates that we can exploit the following CRN to reconstruct the final estimation $\widehat { {\mathbf{H}}}_{{\text{DL}}}^{\text{sf}}$.

\section{Simulation Results}
In this section, we provide numerical results to verify the effectiveness of the proposed MDDL-based channel estimation and feedback scheme.
First, we elaborate the implementation details and parameters adopted in our simulations setting.
Then, since the TDD uplink channel estimation and FDD downlink channel estimation share the same processing mechanism, we take the downlink channel estimation and feedback for FDD systems as examples to evaluate the performance.
Finally, we investigate the performance of the proposed MDDL-based scheme in scenarios with fixed scattering  environments, which is discussed in more detail next.
\subsection{Simulation Setup}
In our simulations, we consider that the BS is equipped with an ULA with ${N_{{\text{BS}}}} = 256$ and ${N_{{\text{RF}}}} = 4$ RF chains.
The number of OFDM subcarriers in the channel estimation phase is set to $K = 64$. The redundant dictionary with an oversampling ratio $G{\text{/}}{N_{{\text{BS}}}}{\text{ = 4}}$ is considered, i.e., the quantized angle grids $G$ is set to $1024$.
In addition, the experiments are performed in PyCharm Community Edition (Python 3.6 environment and Tensorflow 1.13.1) on a computer with dual Intel Xeon 8280 CPU (2.6GHz) and dual Nvidia GeForce GTX 2080Ti GPUs.

The proposed MMV-LAMP network based CRN is composed of $T = 5$ layers, where each layer has the same network structure with the trainable parameters $\left\{ {{\bm{\Xi }}_{\text{DL}},{\mathbf{B}}_{\text{DL}},{\bm{\theta }}_{\text{DL}}} \right\}$.
While the proposed MMV-LAMP based FRSN is composed of $T' = 2$ layers, where the trainable parameters are $\left\{ {{\mathbf{B}'},{\bm{\theta }'}} \right\}$.
For the training of MMV-LAMP network, we generate a training set including ${S_{{\text{tr}}}} = 5000$ spatial-frequency domain channel samples according to the channel model in (\ref{frequency channel}), so that the dimension of the input channel samples is $( {{S_{{\text{tr}}}},{N_{{\text{BS}}}},K} )$.
Similarly, the parameters of the validation set and the test set are ${S_{{\text{va}}}} = 2000$ and ${S_{{\text{te}}}} = 1000$, respectively.
We choose the NMSE as the metric for performance evaluation, which is defined as
\begin{equation}\label{NMSE}
	{\text{NMSE}}( {{\mathbf{H}},\widehat {\mathbf{H}}} ) =  10{\text{log}}_{10}( {{\mathbb{E}}\left[ {\frac{{\left\| {{\mathbf{H}} - {\mathbf{\widehat H}}} \right\|_F^2}}{{\big\| {\mathbf{H}} \big\|_F^2}}} \right]} ).
\end{equation}

\subsection{MDDL-Based FDD Downlink Channel Estimation}
\begin{figure}[t]
	\centering
	\includegraphics[width=244pt]{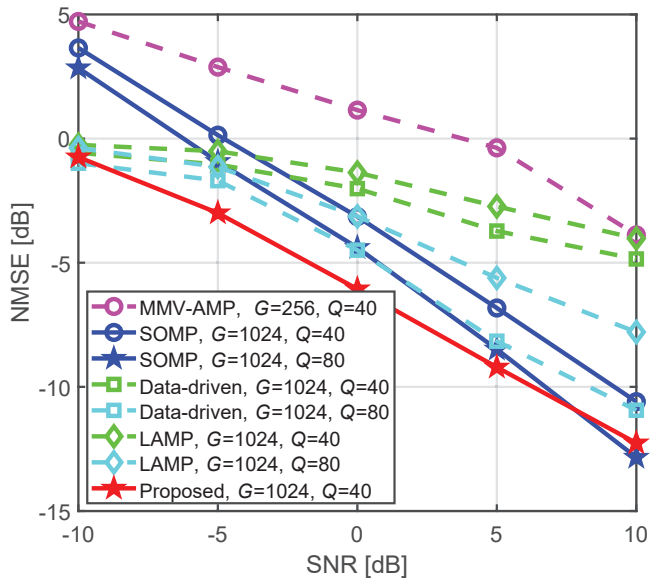}
	\caption{NMSE performance comparison of different channel estimation schemes versus SNRs.}
	\label{pilot}
\end{figure}

As shown in Fig. \ref{pilot} $\footnote{This simulated results for $\{M = 40, Q = 10, N_{\text{RF}} = 4\}$ are equivalent to the uplink channel estimation results for $\{M = 40, Q = 40, N_{\text{RF}} = 1\}$, $\{M = 40, Q = 20, N_{\text{RF}} = 2\}$, and $\{M = 40, Q = 5, N_{\text{RF}} = 8\}$, as long as $M = QN_{\text{RF}}$.}$, we plot the NMSE performance ${\text{NMSE}}( {{\mathbf{H}}_{{\text{DL}}}^{{\text{sf}}},\widehat {\mathbf{H}}_{{\text{DL}}}^{{\text{sf}}}} )$ of the different schemes as a function of signal-to-noise ratio (SNR), including the proposed MDDL-based channel estimation scheme using the MMV-LAMP network, the data-driven deep learning based channel estimation scheme \cite{MXS_TVT}, the LAMP-based channel estimation scheme \cite{LAMP_TSP},
and two model-based channel estimation schemes using the MMV-AMP algorithm \cite{KML_TSP} and the SOMP algorithm \cite{GZ_WC2}.
The number of propagation paths is $L = 8$.
Note that the AMP-based channel estimation scheme requires the measurement matrix's elements to be independent identically distributed, so we don't consider the redundant dictionary matrix (i.e., $G = {N_{{\text{BS}}}} = 256$).

We can observe that the proposed channel estimation scheme outperforms the other channel estimation schemes even with a smaller pilot overhead.
This observation indicates that the proposed channel estimation scheme can achieve better NMSE performance while keeping the pilot overhead to a low level.
This is because the network architecture (i.e., the cascaded CCN and MMV-LAMP-based CRN) of the MDDL-based approach is constructed based on known physical mechanisms and some {\it{a priori}} model knowledge, which can reduce the number of trainable parameters to be learned and can fully take advantage of both model-based algorithms and deep learning methods.
We also observe that the proposed channel estimation scheme can significantly improve the NMSE performance in the low SNR regime compared with other channel estimation schemes.
Therefore, the proposed scheme can reliably reconstruct the high-dimensional channel with a much reduced pilot overhead.

To clearly present the percentage of the reduced pilot overhead compared to state-of-the-art algorithms, we compare the proposed scheme with four state-of-the-art algorithms at SNR=0dB and SNR=5dB, as shown in Table \ref{pilotoverhead}.
We can observe that the proposed scheme with $Q=40$ outperforms the MMV-AMP algorithm, the LAMP network, and data driven deep learning methods with $Q=80$ in the whole range of SNR.
Therefore, we conclude that the proposed scheme can reduce the pilot overhead by at least $50\%$ while achieving the same or even better channel estimation NMSE performance.
\begin{figure}[t]
	\centering
	\includegraphics[width=244pt]{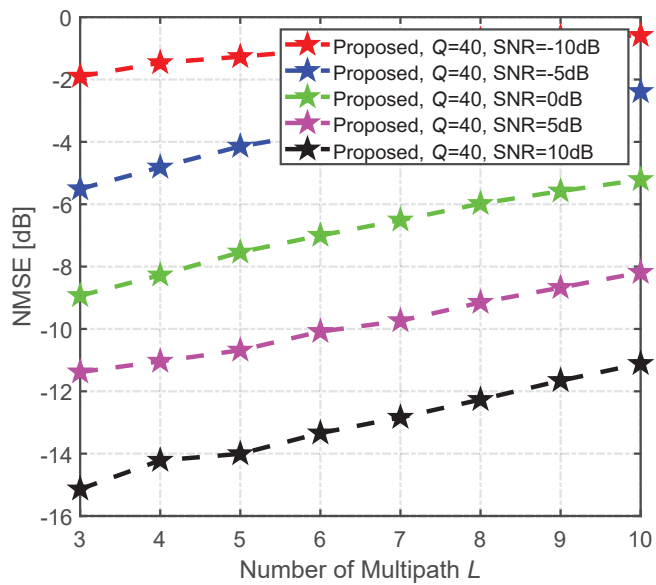}
	\caption{NMSE performance comparison of the proposed scheme versus the number of multipath $L$.}
	\label{path}
\end{figure}
\renewcommand\thetable{\Roman{table}}
\begin{table}[t]
\centering
\renewcommand\arraystretch{1.5}
\caption{NMSE in {\upshape dB}}\label{pilotoverhead}
\begin{tabular}{c|c|c|c|c}
\Xhline{1.2pt}
\multirow{2}*{Channel Estimation Schemes} & \multicolumn{2}{c}{SNR=0dB} & \multicolumn{2}{|c}{SNR=5dB} \\
\cline{2-5}
 & {{\it Q}=40} & {{\it Q}=80} & {{\it Q}=40} & {{\it Q}=80}\\
\Xhline{1.2pt}
MMV-AMP & 1.15 & -0.31 & -0.37 & -7.62 \\
 LAMP & -1.37 & -3.13 & -2.73 & -5.61 \\
Data Driven Deep Learning & -2.02 & -4.49 & -3.72 & -8.16 \\
SOMP & -3.14 & -4.39 & -6.82 & -8.48  \\
Proposed & {\textbf{-6.06}} & & {\textbf{-9.21}} &  \\
\Xhline{1.2pt}
\end{tabular}
\end{table}
\begin{figure}[t]
	\centering
	\includegraphics[width=244pt]{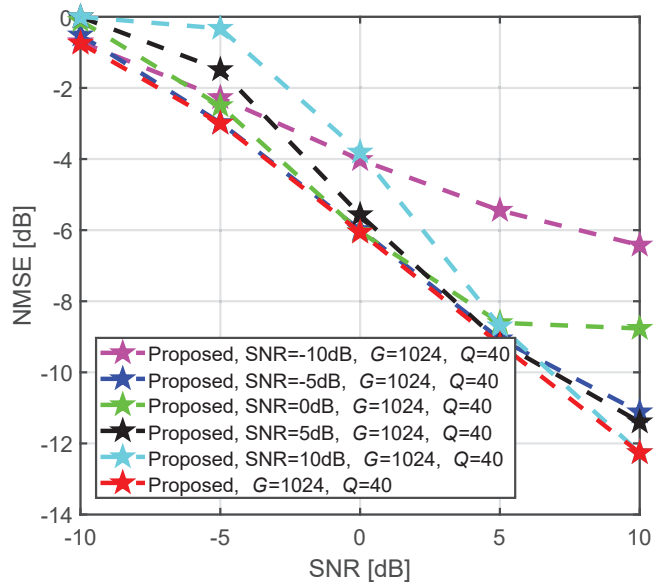}
	\caption{NMSE performance comparison of the proposed scheme trained at the different SNRs.}
	\label{snr}
\end{figure}

We further investigate the robustness of the proposed channel estimation scheme as a function of the number of multipath $L$ in Fig. \ref{path}.
Note that the proposed MMV-LAMP based CRN is trained during the offline training stage, which is based on the channel samples with $L = 8$ multipath components.
However, at the online estimation stage, we observe that the proposed scheme can be robustly adopted to estimate multipath channels with $L \ne 8$, without having to retrain the entire network architecture.
In Fig. \ref{snr}, we show the NMSE performance of the proposed scheme when it is trained based for different SNR values. For example, the setup denoted by ``Proposed, SNR=-10dB, $G=1024$, $Q=40$" means that the proposed scheme was trained at the SNR=-10dB, but tested in the whole SNR range, and the setup denoted by ``Proposed, $G=1024$, $Q=40$" means that the proposed scheme was trained and tested for the same values of SNR.
We can observe that the proposed scheme trained at SNR=-5dB is robust in the low SNR regime, and the NMSE performance is deteriorated just at SNR=10dB.
Therefore, the proposed MMV-LAMP based channel estimator enjoys a better robustness and generalization capability to different channel conditions.

\begin{figure}[t]
	\centering
	\includegraphics[width=244pt]{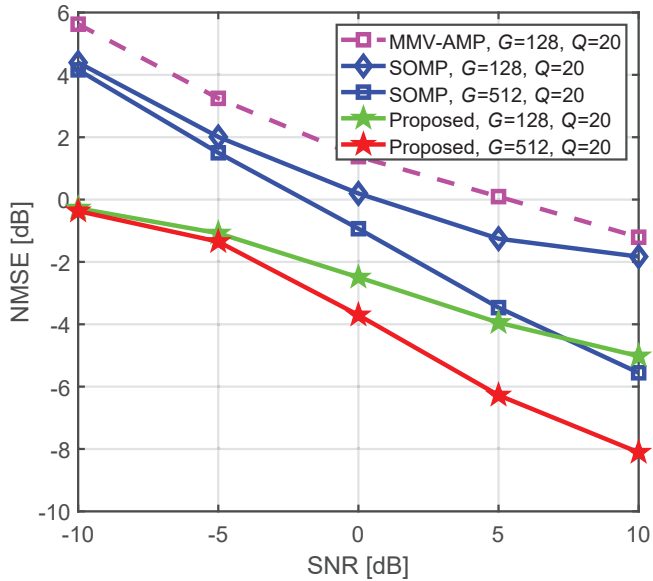}
	\caption{NMSE performance comparison of different channel estimation schemes versus SNRs.}
	\label{128antennas}
\end{figure}
\begin{figure}[t]
	\centering
	\includegraphics[width=244pt]{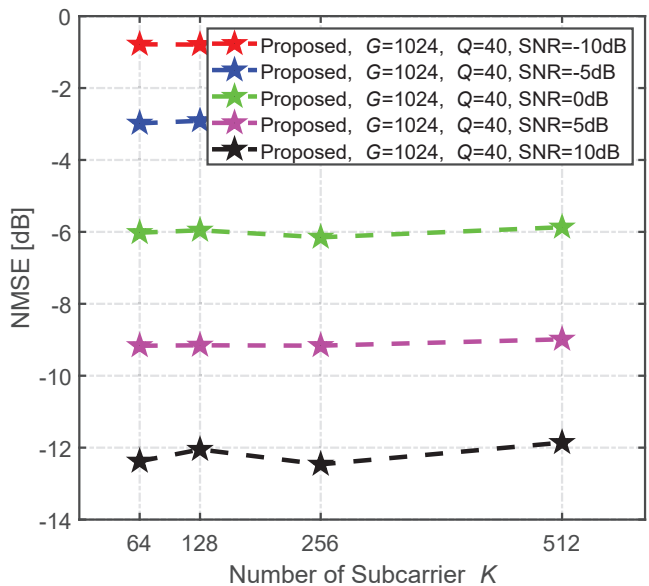}
	\caption{NMSE performance comparison of the proposed scheme versus the number of subcarrier $K$.}
	\label{carriers_number}
\end{figure}
As for the number of antennas $N_{\text{BS}}$, we further investigated the channel estimation NMSE performance with $N_{\text{BS}} = 128$ in Fig. \ref{128antennas}.
Specifically, since the dimension of the beamforming/combining matrix is related to the number of antennas, changing the number of antennas requires retraining the CCN and CRN.
In order to meet the same compression ratio, i.e., $Q/{N_{\text{BS}}}=40/256$, the adopted pilot overhead is $Q = 20$.
We can observe that the proposed channel estimation scheme outperforms the other channel estimation schemes in this case.

As for the different numbers of subcarriers $K$, as shown in Fig.\ref{carriers_number}, we have investigated the channel estimation NMSE performance with $K = 64, 128, 256, 512$.
Specifically, since the proposed channel estimation scheme was trained under the number of subcarriers $K=64$, we divide the subcarriers evenly into multiple sub-groups (each sub-group has 64 subcarriers) to directly apply the trained channel estimation network. 
For instance, for the number of subcarriers $K=128$, we divide the subcarriers into $128/64 = 2$ sub-groups, i.e., the subcarrier indices of each sub-group are respectively $\left\{ {1,3,5, \cdots ,127} \right\}$ and $\left\{ {2,4,6, \cdots ,128} \right\}$.
We can observe that the proposed scheme trained with 64 subcarriers can be effectively applied to the case $K=$ 128, 256, and 512, which further proves the robustness of the proposed scheme.

\begin{figure}[t]
	\centering
	{\subfigure[]{\label{dictionary}
	\centering
	\includegraphics[width=244pt]{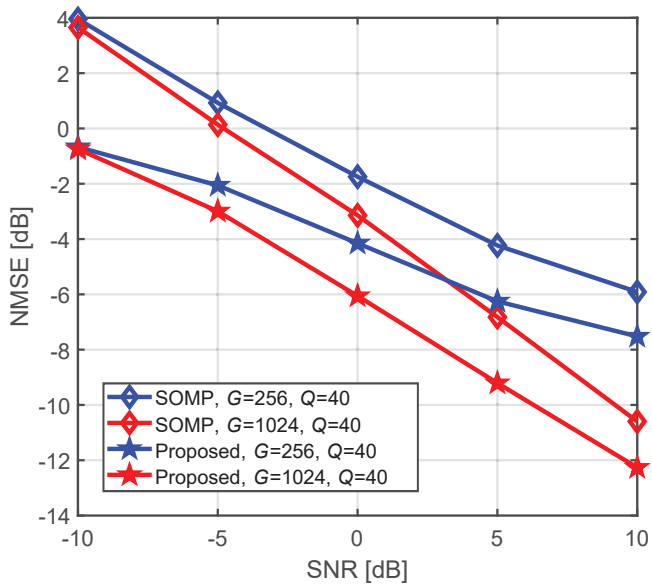}
	}}
	{\subfigure[]{\label{carrier}
	\centering
	\includegraphics[width=244pt]{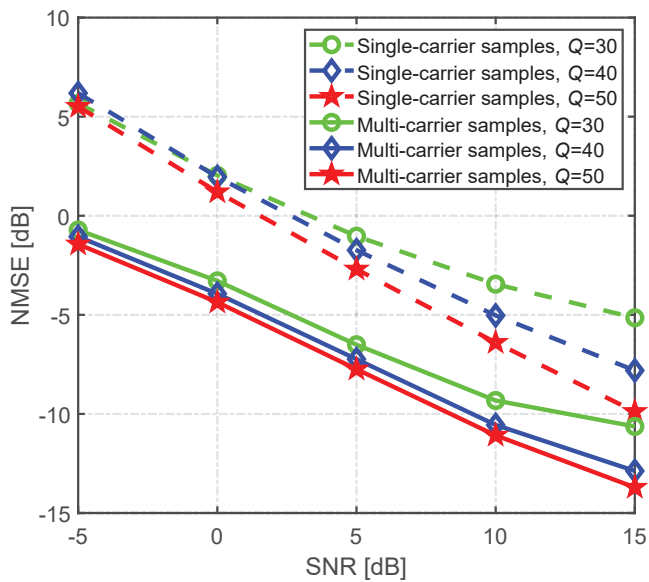}
	}}
	\caption{NMSE performance comparison of the proposed MDDL-based channel estimation scheme versus SNRs, where (a) the effectiveness of the devised redundant dictionary matrix, (b) the effectiveness of the proposed scheme using multi-carrier channel samples to train.}	
\end{figure}

As mentioned above, we design a redundant dictionary matrix to combat the power leakage problem by quantizing the angles with a finer resolution.
Therefore, in Fig. \ref{dictionary},
we also compare the performance of the proposed MMV-LAMP based channel estimator without using redundant dictionary matrix, i.e., $G = {N_{{\text{BS}}}} = 256$, to demonstrate the effectiveness of the redundant dictionary matrix.
We can find that the proposed MMV-LAMP based and the SOMP-based channel estimators can improve the sparse channel estimation performance by utilizing the redundant dictionary matrix to cope with the power leakage problem.

Note that the effectiveness of the proposed MMV-LAMP based channel estimator for OFDM systems is based on a training dataset based on multi-carrier channel samples, which consist of the channels of all subcarriers from different channel realizations.
To verify the training effectiveness of multi-carrier channel samples, we further compare the performance of the proposed scheme based on single carrier channel samples, as shown in Fig. \ref{carrier}.
We can observe that the proposed MMV-LAMP based channel estimator trained by multi-carrier channel samples can exhibit more excellent performance.

We further investigate the computational complexity. In the case of offline training, specifically, the computation complexity is not a major concern, because the required time is usually not strictly limited. 
Moreover. we discuss the computational complexity of the different channel estimation schemes as follows.
\begin{itemize}
\item As for the data driven deep learning based channel estimation scheme, its main steps in testing stage include: i) two fully-connected operations with computational complexity $\mathcal{O}( 2MG )$; ii) $N_{\text{re}}$ convolutional operations with computational complexity $\mathcal{O}( {G{\beta ^2}\sum\limits_{i = 1}^{{N_{{\text{re}}}}} {{n_{i - 1}}{n_i}} } )$, where $\beta$ is the side length of the convolutional filters, $n_{i-1}$ and $n_i$ denote the numbers of input and output feature maps of the $i$-th convolutional layer $(1 \le i \le {N_{{\text{re}}}})$, respectively.
Therefore, the computational complexity of the data driven deep learning based channel estimation is $\mathcal{O}( {2MG + G{\beta ^2}\sum\limits_{i = 1}^{{N_{{\text{re}}}}} {{n_{i - 1}}{n_i}} } )$.
\item The proposed channel estimation scheme is developed from the MMV-AMP algorithm, which mainly requires matrix multiplication operations, Therefore, the MMV-AMP algorithm, the LAMP network, and the proposed MMV-LAMP network share similar computational complexities, i.e., $\mathcal{O}( {M{N_{{\text{BS}}}}K + TMGK} )$ in the case of $K$ subcarriers.
\item As for the SOMP algorithm, we denote the number of iterations as $I$ and its main steps include: i) correlation operation with computational complexity ${\mathcal{O}}(M{G^2}KI)$; ii) project subspace operation with computational complexity ${\mathcal{O}}(\frac{1}{4}{I^2}{(I + 1)^2} + \frac{1}{3}MI(I + 1)(2I + 1) + \frac{1}{2}MKI(I + 1))$; iii) update residual operation with computational complexity ${\mathcal{O}} ( {\frac{1}{2}\rho {N_{{\rm{BS}}}}KI(I + 1)} )$. 
Therefore, the computational complexity of the SOMP algorithm is ${\mathcal{O}}(M{G^2}KI$ $ + \frac{1}{4}{I^2}{(I + 1)^2}$ $+ \frac{1}{3}MI(I + 1)(2I + 1) + \frac{1}{2}MKI(I + 1) + {\frac{1}{2}\rho {N_{{\rm{BS}}}}KI(I + 1)})$.
\end{itemize}
Finally, we provide the computational complexity analysis of the different channel estimation schemes as shown in Table \ref{CC}.
\begin{table}[t]
  \centering
  \renewcommand\arraystretch{2.5}
\caption{Computational Complexity of Different Channel Estimation Schemes}\label{CC}
  \begin{tabular}{c|c}
    \Xhline{1.2pt}
    {\bf{Schemes}} & {\bf{Complexity}} \\
    \Xhline{1.2pt}
    Data driven deep learning & $\mathcal{O}( {2MG + G{\beta ^2}\sum\limits_{i = 1}^{{N_{{\text{re}}}}} {{n_{i - 1}}{n_i}} } )$ \\
    \hline
    SOMP & \makecell[c]{${\mathcal{O}}(M{G^2}KI$ $ + \frac{1}{4}{I^2}{(I + 1)^2}$ \\ $+ \frac{1}{3}MI(I + 1)(2I + 1)$ \\ $+ \frac{1}{2}MKI(I + 1) + {\frac{1}{2}\rho {N_{{\rm{BS}}}}KI(I + 1)})$} \\
    \hline
    LAMP &  $\mathcal{O}( {M{N_{{\text{BS}}}}K + TMGK} )$\\
    \hline
    MMV-AMP & $\mathcal{O}( {M{N_{{\text{BS}}}}K + TMGK} )$ \\
    \hline
    Proposed & $\mathcal{O}( {M{N_{{\text{BS}}}}K + TMGK} )$ \\
    \Xhline{1.2pt}
  \end{tabular}
\end{table}

\subsection{Channel Estimation Under Non-Ideal Hardware Constraints}
\begin{figure}[t]
	\centering
	{\subfigure[]{\label{phase quan}
	\centering
	\includegraphics[width=244pt]{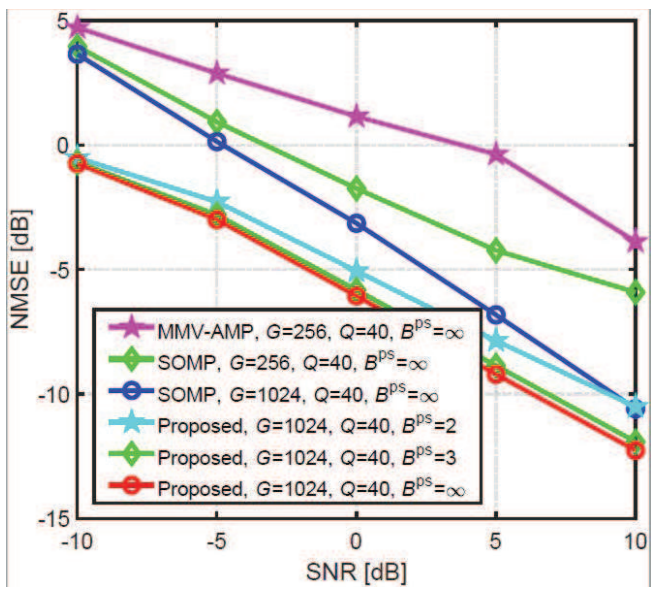}
	}}
	{\subfigure[]{\label{ADC quan}
	\centering
	\includegraphics[width=244pt]{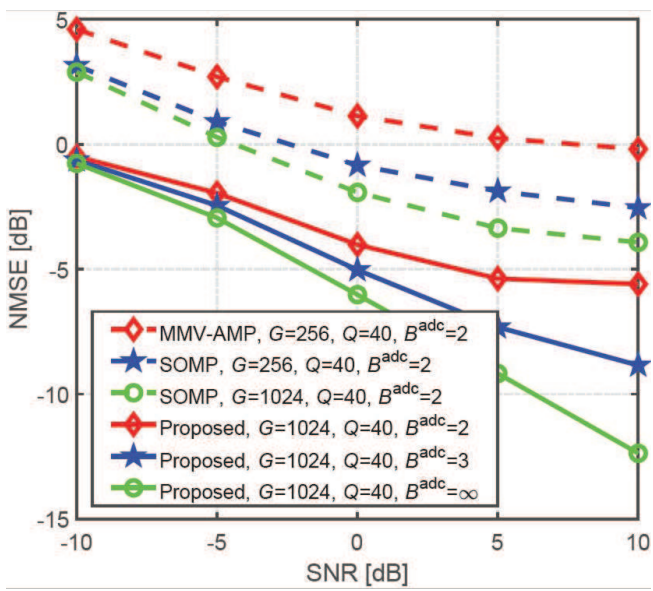}
	}}
	\caption{NMSE performance comparison of the proposed MDDL-based channel estimation scheme versus SNRs, where (a) phase shift quantization, (b) analog-to-digital converter (ADC) quantization.}	
\end{figure}

In practical systems, the phase of each combining and beamforming matrix coefficient is not a continuous value.
Therefore, we further investigate the performance of the proposed MDDL-based channel estimation scheme under the constraint of PSN with a finite phase resolution.
Specifically, after the offline training, the CCN including the continuous complex-valued combining matrix ${\mathbf{F}}_{\text{UL}} = \frac{1}{{\sqrt {{N_{{\text{BS}}}}} }}\exp \left[ {{\text{j}}( {\bm{\Xi }}_{\text{UL}} )} \right]$ in (\ref{W}) and the beamforming matrix ${\mathbf{F}}_{\text{DL}} = \frac{1}{{\sqrt {{N_{{\text{BS}}}}} }}\exp \left[ {{\text{j}}( {\bm{\Xi }}_{\text{DL}} )} \right]$ in (\ref{F}) are quantized to the elements in the set ${\mathbf{\Delta }}$ according to the minimum Euclidean distance criterion. That is to say, after quantization, we have $\left\{ {{\mathbf{\Xi }}_{\text{UL}},{\mathbf{\Xi}}_{\text{DL}}} \right\} \in {\mathbf{\Delta }}$,
and ${\mathbf{\Delta }}$ is the quantized phase set of the PSN whose resolution is ${B^{{\text{ps}}}}$, given by
\begin{equation}\label{phase quan eq}
	{\mathbf{\Delta }} = \left\{ {0,\frac{{2\pi }}{{{2^{{B^{{\text{ps}}}}}}}},2 \cdot \frac{{2\pi }}{{{2^{{B^{{\text{ps}}}}}}}}, \cdots ,2\pi  - \frac{{2\pi }}{{{2^{{B^{{\text{ps}}}}}}}}} \right\}.
\end{equation}
The network architecture and training strategy of the CRN remains unchanged$\footnote{Since the quantized phase shifters will result in non-differential gradients, it is not feasible to directly use the Adam algorithm.
To avoid this challenge, we consider a simple method that the offline training is performed by assuming the infinite precision phase resolution. At the online test stage, the combining/beamforming matrix are quantized according to the minimun Euclidean distance criterion. Note that the authors of \cite{beamforming_ICC} proposed a sub-optimum quantization method to solve this issue, which may be integrated into the proposed scheme for better performance in our future work..}$.
As shown in Fig. \ref{phase quan}, we plot the NMSE performance ${\text{NMSE}}( {{\mathbf{H}}_{{\text{DL}}}^{{\text{sf}}},\widehat {\mathbf{H}}_{{\text{DL}}}^{{\text{sf}}}} )$ of the proposed channel estimation scheme as a function of the SNR, where the resolution of the PSN is set to ${B^{{\text{ps}}}} = 2, 3$ bits.
We observe that the proposed MMV-LAMP based channel estimator works well in the case of 3-bit quantization and suffers from a little loss in the case of 2-bit quantization.
This observation further demonstrates the robustness of the proposed MMV-LAMP based channel estimator under non-ideal PSN with limited resolution.

Moreover, considering the limited resolution of the analog-to-digital converter (ADC) at the BS, the downlink received signals are first quantized by the ADC in the time domain, so the received frequency-domain pilot signals after time-domain quantization can be expressed as
\begin{equation}\label{ADC quan eq}
{ {{\mathbf{Y}}_{{\text{DL}}}^{{\text{quan}}}} } = {{\lambda ^{{\text{quan}}}}{( {\mathbf{Y}}_{{\text{DL}}}{\mathbf{U}} )}}{{{\mathbf{U}}^{\text{H}}}},
\end{equation}
where ${{\mathbf{Y}}_{{\text{DL}}}{\mathbf{U}}}$ and ${{\lambda ^{{\text{quan}}}}{( {\mathbf{Y}}_{{\text{DL}}}{\mathbf{U}} )}}$ are the received time-domain signals before the ADC and after the ADC, respectively,
and $\lambda ^{{\text{quan}}}(  \cdot  )$ is the complex-valued quantization function. This quantization function is applied to the received signals element-wise, and the real and imaginary parts are quantized separately. Here, we consider a uniform codebook for quantization as
\begin{equation}\label{quan function}
	C = \left\{ { - \frac{{{2^{{B^{{\text{adc}}}}}} - 1}}{2}\varepsilon , \cdots ,\frac{{{2^{{B^{{\text{adc}}}}}} - 1}}{2}\varepsilon } \right\},
\end{equation}
where ${B^{{\text{adc}}}}$ is the number of quantization bits, $\varepsilon  = \left[ {y_{\max } - y_{\min }} \right]\big/{2^{{B^{{\text{adc}}}}}}$, $y_{\max} $
and $y_{\min }$ are the maximum and the minimum real value of both the real and imaginary parts of ${\mathbf{Y}}_{{\text{DL}}}{\mathbf{U}}$, respectively.
Specifically, we first train the CCN's parameters ${\left\{ {{{\bm{\Xi }}_{{\text{DL}}}}} \right\}}$ and CRN's parameters ${\left\{ {{{\mathbf{B}}_{{\text{DL}}}},{{\bm{\theta }}_{{\text{DL}}}}} \right\}}$ based on the ADC with infinite resolution,
then we adopt the above quantization method to obtain the quantized frequency-domain signal ${ {{\mathbf{Y}}_{{\text{DL}}}^{{\text{quan}}}} }$.
Finally, we input the quantized ${ {{\mathbf{Y}}_{{\text{DL}}}^{{\text{quan}}}} }$ directly into the previously trained CRN to reconstruct the channel matrix ${{\widehat{{\mathbf{H}}}}_{\text{DL}}^{{\text{sf}}}}$.

As shown in Fig. \ref{ADC quan}, we plot the NMSE performance ${\text{NMSE}}( {{\mathbf{H}}_{{\text{DL}}}^{{\text{sf}}},\widehat {\mathbf{H}}_{{\text{DL}}}^{{\text{sf}}}} )$ of the proposed channel estimation scheme as a function of the SNR, where the number of quantization bits is set to ${B^{{\text{adc}}}} = 2, 3$.
We observe that the performance of the MMV-AMP-based and the SOMP-based channel estimation schemes degrade.
However, the proposed channel estimation scheme can still work even in the low SNR regime. Considering that the SNR is usually low in most mmWave systems at the channel estimation stage, the proposed scheme is effective in estimating the channels with non-ideal ADC at the receiver.
\begin{figure}[t]
	\centering
	\includegraphics[width=252pt]{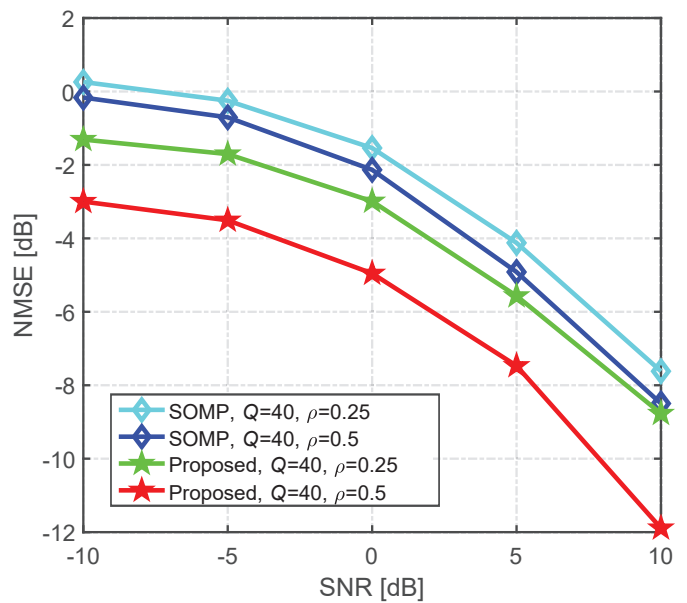}
	\caption{Channel reconstruction NMSE performance comparison of the proposed MDDL-based channel feedback scheme versus SNRs.}
	\label{CS_radio}
\end{figure}

\subsection{MDDL-Based FDD Downlink Channel Feedback}
In this section, we investigate the channel reconstruction performance ${\text{NMSE}}( {{\mathbf{H}}_{{\text{DL}}}^{{\text{sf}}},\widehat {\mathbf{H}}_{{\text{DL}}}^{{\text{sf}}}} )$ of the proposed channel feedback scheme.
Specifically, as shown in Fig. \ref{dimension reduction network}, the CCN at the BS is the same as that used at the downlink channel estimation stage,
then the noisy frequency-domain received signal ${\mathbf{Y}}_{{\text{DL}}}^{\text{T}}$ is obtained at the user.
Moreover, in order to compress the feedback overhead, the user only feeds ${\mathbf{Y}}_{{\text{DL}}}^{\text{T}}$ on $K_c$ of $K$ subcarriers back to the BS.
Finally, the BS exploits the proposed MMV-LAMP based FRSN and the CRN to recover the spatial-frequency domain channel matrix $\widehat {\mathbf{H}}_{{\text{DL}}}^{{\text{sf}}}$.
In addition, we define the feedback compression radio as $\rho {\text{ = }}{K_c}/K$.
As shown in Fig. \ref{CS_radio}, we can observe that the proposed MMV-LAMP based FCRN (including the FRSN and the following CRN) with $\rho = 0.25$, i.e., $K_c = 16$, even outperforms the SOMP-based channel feedback scheme with $\rho = 0.5$, i.e., $K_c = 32$.
Therefore, the effectiveness of the proposed MDDL-based channel feedback scheme is verified.

\subsection{Channel Estimation Based on Fixed Scattering Environments}
\begin{figure*}[t]
	\centering
	\includegraphics[scale=0.4]{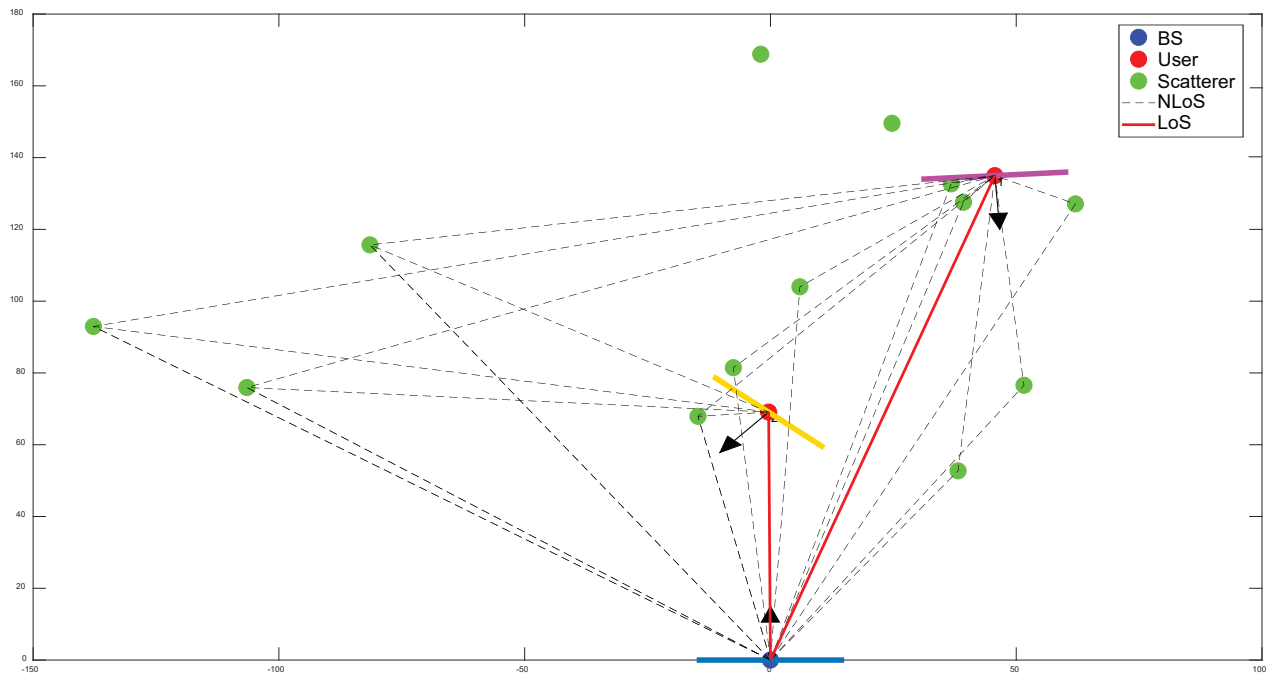}
	\caption{The schematic diagram of the fixed scattering environment.}
	\label{envir_channel}
\end{figure*}

In this section, the TDD uplink channel estimation NMSE performance ${\text{NMSE}}( {{\mathbf{H}}_{{\text{UL}}}^{{\text{sf}}},\widehat {\mathbf{H}}_{{\text{UL}}}^{{\text{sf}}}} )$ is investigated.
To evaluate the superiority of the proposed learning strategy that jointly trains the pilots and channel estimator, we consider the scenario with fixed scattering environments. 
The fixed scattering environment is shown in Fig. \ref{envir_channel}, in which the positions of the BS, the user, and the scatterers are marked by blue, red, and green, respectively.
The red solid line represents the line-of-sight (LoS) link between the BS and the user, and the black dotted line indicates the non-line-of-sight (NLoS) link via the scatterer.
Note that for the sake of simplifying the channel generation, we only consider a single-bounce NLoS channel model.
As for the array setting, we take the BS side as an example, i.e., the bold blue solid line represents the antenna array of the BS, whose normal direction is marked as the black arrow.
The antenna array setting at users is the same as the BS.

Next, we describe how to generate the channel samples based on the fixed scattering environment.
Specifically, as shown in Fig. \ref{envir_channel}, the scenario consists of a BS and a number of users geographically distributed in a certain outdoor environment, in which the scatterers are randomly distributed. 
For the given fixed scattering environment, we can generate the channel samples based on the channel parameters (including the AoAs/AoDs, path loss), which can be calculated based on the geometric characteristics between the BS and the user.
More specifically,
\begin{itemize}
\item \textbf{AoAs/AoDs:} 
As for the uplink channel estimation, the AoA ${\phi _{{\text{BS}}}}$ at the BS is the angle relative to the horizontal axis.
For the user, the AoD ${\phi _{{\text{UE}}}}$ at the user follows the uniform distribution $\left[ {0,2\pi } \right)$, which is defined as the angle between the normal direction of the user array and the horizontal axis.
\item \textbf{Path loss:}
Taking the NLoS link as an example, the large-scale fading gain $G_l$ can be modeled based on the free-space path loss of Friis' formula as
\begin{equation}
{G_l} = 20{\log _{10}}( {\frac{{4\pi d_{l,1}}}{{{\lambda _c}}}} ) + 20{\log _{10}}( {\frac{{4\pi d_{l,2}}}{{{\lambda _c}}}} ) + {G_s},
\end{equation}
where $d_{l,1}$ ($d_{l,2}$) denotes the communication distance between the user and the $l$-th scatterer (the $l$-th scatterer and the BS), ${\lambda _c}$ is the carrier wavelength, and ${G_s}$ denotes the path loss via the $l$-th scatterer.
\end{itemize}

Based on the fixed scattering environment discussed above,
we generate 10000 channel samples, named as fixed scattering environment (FSE) data set, by considering a randomly distributed user location and normal direction of the user array.
In simulations, we divide the 10000 channel samples into 8000, 1000, and 1000, corresponding to the training, validation, and test sets, respectively.
\begin{figure}[t]
	\centering
	\includegraphics[width=252pt]{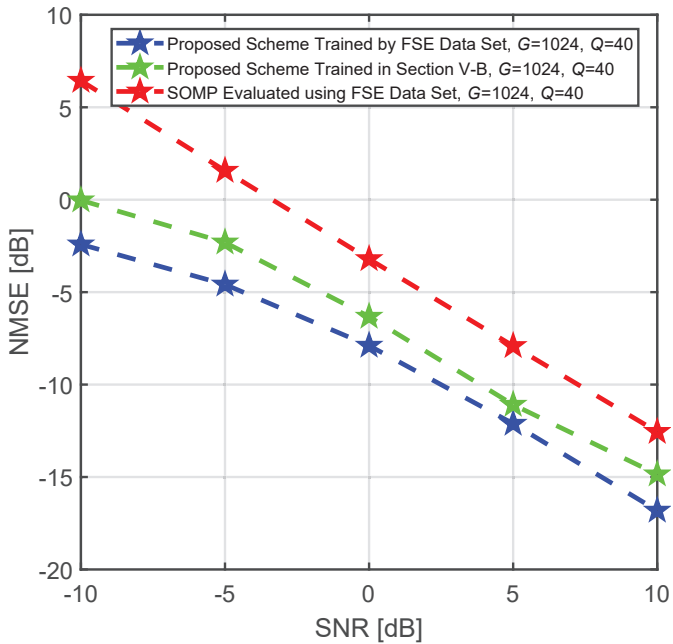}
	\caption{NMSE performance comparison of different channel estimation schemes versus SNRs.}
	\label{Environment}
\end{figure}

\begin{figure}[t]
	\centering
	\includegraphics[width=252pt]{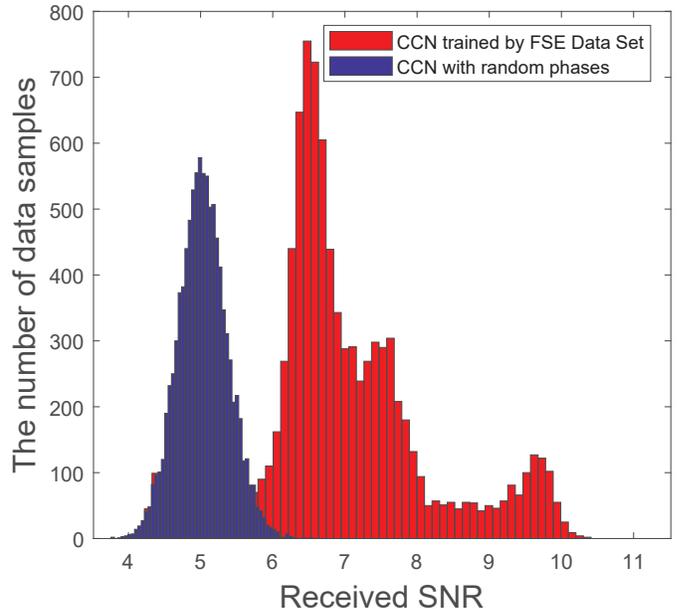}
	\caption{The received SNR distributions at the user side using different CCNs.}
	\label{E1}
\end{figure}

As shown in Fig. \ref{Environment}, we plot the uplink channel estimation performance ${\text{NMSE}}( {{\mathbf{H}}_{{\text{UL}}}^{{\text{sf}}},\widehat {\mathbf{H}}_{{\text{UL}}}^{{\text{sf}}}} )$ of the different schemes as a function of the SNR, including the proposed scheme trained by using the FSE data set, the proposed scheme trained in Section V-B (i.e., not trained with the FSE data set), and the SOMP algorithm using random combining matrix, where we consider $G=1024$, $Q=40$.
Note that, at the NMSE performance evaluation phase, the input channel samples for these schemes come from the test set of the FSE data set.  
We can observe that the proposed channel estimation scheme (including the combining matrix ${\mathbf{F}}_{{\text{UL}}}^{\text{H}}$ and the MMV-LAMP network based channel estimator) trained with FSE data set can effectively learn the channel environment characteristics using less pilot overhead.
Moreover, we can utilize the trained combining matrix ${\mathbf{F}}_{{\text{UL}}}^{\text{H}}$ in the uplink as the beamforming ${\mathbf{F}}_{{\text{DL}}}^{\text{T}}$ in the downlink channel estimation for improving the receive SNR at the users as verified in Fig. \ref{E1}.

Fig. \ref{E1} depicts the received SNR distributions at the users using two different CCN designs in the downlink channel estimation phase.
Specifically, ``CCN trained by FSE Data Set" indicates that the BS adopts the beamforming matrix ${\mathbf{F}}_{{\text{DL}}}^{\text{T}}$ or CCN being the same as the combining matrix ${\mathbf{F}}_{{\text{UL}}}^{\text{H}}$ trained in Fig. \ref{Environment}, while ``CCN with random phases" indicates that the phases of the beamforming matrix ${\mathbf{F}}_{{\text{DL}}}^{\text{T}}$ adopted by the BS are random.
In Fig. \ref{E1}, under the fixed scattering environment, we can observe that more users can achieve a larger received SNR by exploiting the proposed CCN trained with the FSE data set  than the CCN with random phases.

\section{Conclusions}
In this paper, we have proposed an MDDL-based channel estimation and feedback scheme for wideband mmWave massive hybrid MIMO systems, where the angle-delay domain channels' sparsity is exploited for reduced overhead.
First, we have considered the uplink channel estimation for TDD systems.
To reduce the uplink pilot overhead for estimating the high-dimensional channels from a limited number of RF chains at the BS,
we have proposed to jointly train the PSN and the channel estimator as an auto-encoder.
Specifically, by exploiting channels' structured sparsity from an {\it{a priori}} model and learning the integrated trainable parameters from the data samples, the MMV-LAMP network with the devised redundant dictionary has been proposed to jointly recover multiple subcarriers' channels with significantly enhanced performance. 
Moreover, we have considered the downlink channel estimation and feedback for FDD systems.
Similarly, the pilots at the BS and channel estimator at the users can be jointly trained as an encoder and a decoder, respectively.
To further reduce the channel feedback overhead, only the received pilots on part of the subcarriers are fed back to the BS, which can exploit the proposed MMV-LAMP network to reconstruct the spatial-frequency channel matrix.
Further, we consider to generate the data samples from the scenarios with fixed scattering environments, and optimize the combining/beamforming matrix and MMV-LAMP network by learning and perceiving the characteristics of the channel environments for the improved performance.
Simulation results have verified that the proposed MDDL-based solution can achieve a significant improvement in channel estimation and feedback performance than the conventional schemes.

\appendix
First, we introduce the AMP algorithm.
We consider a typical single-measurement-vector (SMV) CS problem
\setcounter{equation}{37}
\begin{equation}\label{SMV_1_appendix}
{\mathbf{y}} = {\mathbf{Ax}} + {\mathbf{n}},
\end{equation}
where ${\mathbf{y}}{ \in ^{M \times 1}}$ is a noisy measurement, 
${\mathbf{A}} \in {\mathbb{C}^{M \times N}}$ represents the measurement matrix with $M \ll N$, ${\mathbf{x}} \in {\mathbb{C}^{N \times 1}}$ denotes the sparse vector, and ${\mathbf{n}} \in {\mathbb{C}^{M \times 1}}$ is the AWGN.
To reconstruct the sparse vector $\mathbf{x}$, the AMP algorithm can be given as (see equations (10)-(12) in \cite{LAMP_TSP})
\begin{subequations}
	\begin{align}
		{{\mathbf{r}}_t} &= {{\mathbf{\hat x}}_{t - 1}} + {{\mathbf{A}}^{\text{H}}}{{\mathbf{v}}_{t - 1}} , \\
		{{{\mathbf{\hat x}}}_t} &= \eta \left( {{{\mathbf{r}}_t};{{\bm{\theta }}_t},{\sigma _t}} \right), \\
		{{\mathbf{v}}_t} &= {\mathbf{y}} - {\mathbf{A}}{{{\mathbf{\hat x}}}_t} + {b_t}{{\mathbf{v}}_{t - 1}},
	\end{align}
\end{subequations}
where ${{\mathbf{v}}_0} = \mathbf{y}$, ${\mathbf{\hat x}}_0 = \mathbf{0}$. 
And
\begin{align}
	{\sigma _t} &= \frac{1}{{\sqrt {M} }}{\left\| {{{\mathbf{v}}_{t - 1}}} \right\|_2},  \\
	{b_t} &= \frac{1}{M}\sum\limits_{j = 1}^N {\frac{{\partial {{\left[ {\eta \left( {{{\mathbf{r}}_t};{{\bm{\theta }}_t},{\sigma _t}} \right)} \right]}_j}}}{{\partial {r_j}}}},
\end{align}
where ${\left[ {\eta \left( {{{\mathbf{r}}_t};{{\mathbf{\theta }}_t},{\sigma _t}} \right)} \right]_j} = \eta \left( {{{\left[ {{{\mathbf{r}}_t}} \right]}_j};{{\mathbf{\theta }}_t},{\sigma _t}} \right)$, for $1 \le j \le N$.
Moreover, we consider the shrinkage function $\eta \left( { \cdot ; \cdot } \right)$ that corresponds to MSE-optimal denoisers under zreo-mean Bernoulli-Gaussian (BG) priors. That is, $\hat x = {\text{E}}\left\{ {\left. x \right|r} \right\}$, where $x$ has the BG prior
\begin{equation}\label{BG__appendix}
p\left( {x;\gamma ,\phi } \right) = \left( {1 - \gamma } \right)\delta \left( x \right) + \gamma \mathcal{N}\left( {x;0,\phi } \right),
\end{equation}
and $r$ is an AWGN-corrupted measurement of $x$, i.e.,
\begin{equation}\label{r__appendix}
r = x + e{\text{ for }}e \sim \mathcal{N}\left( {0,{\sigma ^2}} \right).
\end{equation}
The MSE-optimal denoiser is then (see equation (39) in \cite{LAMP_TSP})
\begin{equation}\label{x_hat_appendix}
\hat x = \frac{r}{{( {1 + \frac{{{\sigma ^2}}}{\phi }} )( {1 + \frac{{1 - \gamma }}{\gamma }\frac{{\mathcal{N}\left( {r;0,{\sigma ^2}} \right)}}{{\mathcal{N}\left( {r;0,{\sigma ^2} + \phi } \right)}}} )}}.
\end{equation}
Then, to turn (\ref{x_hat_appendix}) into a learnable shrinkage function, i.e., the LAMP algorithm \cite{LAMP_TSP}, we set ${\theta _1} = \phi $ and ${\theta _2} = \log \frac{{1 - \gamma }}{\gamma }$ and then simplify (\ref{x_hat_appendix}) as  (see equation (40) in \cite{LAMP_TSP})
\begin{equation}\label{lamp_appendix}
{[ {\eta ( {{\mathbf{r}};{\bm{\theta }},\sigma } )} ]_j} = \frac{{{r_j}}}{{( {1 + \frac{{{\sigma ^2}}}{{{\theta _1}}}} )( {1 + \sqrt {1 + \frac{{{\theta _1}}}{{{\sigma ^2}}}} \exp [ {{\theta _2} - \frac{{r_j^2}}{{2{\sigma ^2}( {1 + \frac{{{\sigma ^2}}}{{{\theta _1}}}} )}}} ]} )}},
\end{equation}
where the trainable parameters include ${{{\theta _1}}}$ and ${{{\theta _2}}}$, i.e., ${\bm{\theta }} = \left[ {{\theta _1},{\theta _2}} \right]$.

Finnaly, for the developed MMV-LAMP algorithm, we consider a typical MMV CS problem
\begin{equation}\label{MMV_problem_appendix}
	{\mathbf{Y}} = {\mathbf{AX}} + {\mathbf{N}},
\end{equation}
where ${\mathbf{Y}} \in {\mathbb{C}^{M \times K}}$ is a noisy measurement,
${\mathbf{A}} \in {\mathbb{C}^{M \times N}}$ is a measurement matrix,
${\mathbf{X}}\in {\mathbb{C}^{N \times K}}$ denotes the sparse matrix satisfying that its columns $\big\{ {{\mathbf{X}}\left( {:,i} \right)} \big\}_{i = 1}^K$ share the common sparsity, and ${\mathbf{N}} \in {\mathbb{C}^{M \times K}}$ is the AWGN. 
According to \cite{YW_TSP}, we developed the MMV-LAMP algorithm as
\begin{subequations}
	\begin{align}
		{{\mathbf{R}}_t} &= {\mathbf{\widehat X}}_{t - 1} + {{\mathbf{B}}}{{\mathbf{V}}_{t-1}} , \\
		{\mathbf{\widehat X}}_t &= {{\eta }}\left( {{{\mathbf{R}}_t};{{\bm{\theta }}},{\sigma _t}} \right), \\
		{{\mathbf{V}}_t} &= {\mathbf{Y}} - {\mathbf{A\widehat X}}_{t} + {b_t}{{\mathbf{V}}_{t - 1}}, \label{residual_appendix}
	\end{align}
\end{subequations}
where ${{\mathbf{V}}_0} = \mathbf{Y}$, ${\mathbf{\widehat X}}_0 = \mathbf{0}$, and
\begin{align}
	{\sigma _t} &= \frac{1}{{\sqrt {MK} }}{\left\| {{{\mathbf{V}}_{t - 1}}} \right\|_F}, \label{epsilon_t_appendix} \\
	{b_t}{\mathbf{I}} &= \frac{1}{M}\sum\limits_{j = 1}^N {\frac{{\partial {{\left[ {{{\eta}} \left( {{{\mathbf{R}}_t};{{\bm{\theta}}},{\sigma _t}} \right)} \right]}_j}}}{\partial \left[ {{{\mathbf{R}}_t}\left( {j,:} \right)} \right]}}. \label{b_t_appendix}	
\end{align}
Similarly, we consider the zero-mean BG prior and the MSE-optimal denoiser is then (see equations (24)-(26) in \cite{YW_TSP})
\begin{equation}\label{X_hat_appendix}
{\mathbf{\hat x}} = \frac{{\mathbf{r}}}{{( {1 + \frac{{{\sigma ^2}}}{\phi }} )( {1 + \frac{{1 - \gamma }}{\gamma }\frac{{\mathcal{N}\left( {{\mathbf{r}};0,{\sigma ^2}{\mathbf{I}}_{K}} \right)}}{{\mathcal{N}\left( {{\mathbf{r}};0,\left( {{\sigma ^2} + \phi } \right){\mathbf{I}}_{K}} \right)}}} )}},
\end{equation}
where ${\mathbf{\hat x}} \in {\mathbb{C}^{K \times 1}}$. 
We set ${\theta _1} = \phi $ and ${\theta _2} = \log \frac{{1 - \gamma }}{\gamma }$ and then the corresponding shrinkage function $\eta \left( { \cdot ; \cdot } \right)$ can be expressed as
\begin{equation}\label{eta_appendix}
	{[ {\eta ( {{{\mathbf{R}}_t};{{\bm{\theta }}},{\sigma _t}} )} ]_j} = \frac{{{{\mathbf{r}}_{t,j}}}}{{{\pi _t}[ {1 + \exp ( {{\psi _t} - \frac{{{\mathbf{r}}_{t,j}^{\text{H}}{{\mathbf{r}}_{t,j}}}}{{2\sigma _t^2{\pi _t}}}})} ]}},
\end{equation}
where ${{\mathbf{r}}_{t,j}} = {{{\mathbf{R}}_t}( {j,:} )}$ denotes the $j$-th row of the ${{{\mathbf{R}}_t}}$, ${\pi _t}$ and ${\psi _t}$ are respectively given by
\begin{align}
	{\pi _t} &= 1 + \frac{{\sigma _t^2}}{{{\theta _{1}}}}, \label{pi_appendix} \\
	{\psi _t} &= K\log ( {1 + \frac{{{\theta _{1}}}}{{\sigma _t^2}}} ) + {\theta _{2}}. \label{psi_appendix}
\end{align}
The trainable parameters of the developed MMV-LAMP network include $\mathbf{B}$ and $\bm{\theta}$, where ${\bm{\theta }} = \left[ {{\theta _1},{\theta _2}} \right]$.

\end{document}